\documentclass[a4paper,11pt]{article}
\pdfoutput=1 %
\usepackage{jcappub}
\usepackage{rotating}
\usepackage{array}
\usepackage{amsmath,bm}
\usepackage[normalem]{ulem}
\usepackage{slashed}
\usepackage{booktabs}
\usepackage[pdftex,table]{xcolor}
\usepackage{units}
\usepackage{subcaption}
\usepackage{caption,ulem}
\usepackage[T1]{fontenc} 
\usepackage{array}
\usepackage{amsmath,bm}
\usepackage{sectsty}
\allsectionsfont{\boldmath}

\usepackage{tabularx}

\newcolumntype{Y}{>{\centering\arraybackslash}X}

\DeclareUnicodeCharacter{2212}{-}

\arxivnumber{TTK-21-27}

\title{\boldmath Constraining positron emission from pulsar populations with AMS-02 data}

\author[a,b]{Luca Orusa,}
\author[c]{Silvia Manconi,}
\author[a,b]{Fiorenza Donato}
\author[b]{and Mattia Di Mauro}

\affiliation[a]{Dipartimento di Fisica, Università di Torino,\\
  Via P. Giuria 1, Torino, Italy}

\affiliation[b]{Istituto Nazionale di Fisica Nucleare, Sezione di Torino,\\
Via P. Giuria 1, 10125 Torino, Italy}

\affiliation[c]{Institute for Theoretical Particle Physics and Cosmology, RWTH Aachen University, Sommerfeldstr.\ 16, 52056 Aachen, Germany}

\emailAdd{luca.orusa@edu.unito.it}
\emailAdd{manconi@physik.rwth-aachen.de}
\emailAdd{donato@to.infn.it}
\emailAdd{mattia.dimauro@to.infn.it}

\abstract{The cosmic-ray flux of positrons is measured with high precision by the space-borne particle spectrometer AMS-02. The hypothesis that pulsar wind nebulae (PWNe) can significantly contribute to the excess of the positron ($e^+$) cosmic-ray flux has been consolidated after the observation of a $\gamma$-ray emission at TeV energies of a few degree size around Geminga and Monogem PWNe. 
In this work we undertake massive simulations of Galactic pulsars populations, adopting different distributions for their position in the Galaxy,  intrinsic physical properties, pair emission models, in order to overcome the
incompleteness of the ATNF catalog. We fit the $e^+$ AMS-02 data together with a secondary component due to collisions of primary cosmic rays with the interstellar medium. 
We find that several mock galaxies have a pulsar population able to explain the observed $e^+$ flux, typically by few, bright sources. We determine the physical parameters of the pulsars dominating the $e^+$ flux, and assess the impact of different assumptions on radial distributions, spin-down properties, Galactic propagation scenarios and $e^+$ emission time. }

\begin{document}
\maketitle
\flushbottom
\section{Introduction}

The observation of high-energetic cosmic-ray (CR) positrons ($e^+$) first by PAMELA \cite{Adriani:2013uda} and Fermi-LAT \cite{Ackermann_2012} and then with unprecedented precision by AMS-02 \cite{PhysRevLett.122.041102} suggests the presence of primary $e^+$ sources in our Galaxy, as the observed flux exceeds the so-called secondary flux produced by inelastic collisions of CR nuclei in the interstellar medium (ISM) above about 10~GeV, see e.g. \cite{Delahaye:2008ua,Diesing:2020jtm}. Given the intense radiative losses suffered by high energetic electrons and positrons ($e^\pm$) while propagating in the Galaxy, AMS-02 observations require this primary source to be local, i.e.~located within a few kpc from the Earth.
Among the investigated explanations \cite{Mlyshev:2009twa,2009_Grasso,2009_Hooper,2010_Kawanaka,2011_Kashiyama,2012_Kisaka,Gaggero:2013nfa,2013_cholis,2014JCAP...04..006D,Boudaud:2014dta,Tomassetti:2015cva,Lipari:2018usj,Cholis_2018,2016JCAP...05..031D,Mertsch:2020dcy}, pulsars have been consolidating as significant factories of high-energy CR $e^\pm$ in the Galaxy, and thus as main candidates to explain the $e^+$ excess.

First, the pulsar spin-down mechanism  effectively 
produces $e^\pm$ pairs, which are possibly accelerated to multi-TeV energies at the termination shock between the relativistic wind and the surrounding medium (see \cite{Bykov:2017xpo,Amato:2020zfv} for recent reviews on the acceleration mechanism). 
Secondly, observations of pulsars and their surrounding environment, including their pulsar wind nebula (PWN), across the entire electromagnetic spectrum reveal $e^\pm$ accelerated at very-high-energies and generating photons through synchrotron and inverse Compton scattering (ICS) processes \cite{Gaensler_2003,2017hsn..book.2159S}.
In particular, the observation of gamma-ray halos at TeV energies of a few degree size around two nearby pulsars, Geminga and Monogem, reported by Milagro \cite{2009ApJ...700L.127A} and HAWC  \cite{Abeysekara:2017science}  further corroborates the presence of $e^\pm$ accelerated, then escaped, by their PWNe at few tens of parsec away from the pulsar location. 
The observed emission is interpreted as coming from CR $e^\pm$ escaping from the PWN system and ICS low-energy photons of the interstellar radiation field (ISRF). 
Similar extended gamma-ray halos surrounding pulsars have been recently observed by HAWC, by LHAASO at even higher energies \cite{HAWC:2019tcx,LHAASO:2021crt}, and also by {\it Fermi}-LAT at tens of GeV implying even larger angular scales \cite{DiMauro:2019yvh}, suggesting that they might be a general feature of Galactic pulsars \cite{Linden:2017vvb,DiMauro:2019hwn,DiMauro:2020jbz}. While the mechanism shaping $e^\pm$ transport  in these halos, and consequently the gamma-ray emission, is not yet fully understood \cite{Evoli:2018aza,Lopez-Coto:2017pbk,Liu:2019zyj,Fang:2019iym,Recchia:2021kty}, the observed extension is well above the typical PWN size inferred at radio and X-ray wavelengths \cite{DiMauro:2019hwn}. 

Finally, several independent works have demonstrated that pulsar models can provide a good description of AMS-02 $e^+$  and $e^-$  data. 
This conclusion has been reached both by considering the contribution of few nearby sources, as well as the cumulative emission from pulsars as observed in existing catalogs \cite{Mlyshev:2009twa,Boudaud:2014dta,2014JCAP...04..006D,Manconi:2016byt,Fornieri_2020,DiMauro:2020cbn} or in simulations \cite{Cholis_2018,Evoli_2021}, and 
including or not possible effects from suppressed diffusion around sources \cite{Manconi:2020ipm}.
While energy losses limit the distance traveled by  high-energy $e^\pm$ to few hundreds of parsecs, where we expect few Galactic sources contributing significantly, current source catalogs might be not complete. Previous computations, such as \cite{2014JCAP...04..006D}, calculating the contribution of $e^+$ from the ATNF catalog sources 
could therefore suffer for underestimation due to the incompleteness of the catalog. Simulations of the Galactic source population of pulsars are needed to extensively test the pulsar interpretation of the observed $e^+$ flux in order to overcome the limitations of previous studies.

The details of the $e^\pm$ production, acceleration and release from pulsars and their PWN are yet not fully understood \cite{Amato:2020zfv}, as well as  the spatial and energetic distribution of pulsar \cite{Lorimer_2006,Chakraborty:2020lbu}. 
CR $e^+$  are indeed a golden channel to study  primary sources, and in particular PWNe. The CR $e^-$ are
mainly produced by supernova remnants (SNRs), while the pulsar contribution is predicted subdominant \cite{Evoli_2021,DiMauro:2020cbn}. 
High-precision $e^+$ data can now be used to constrain the main properties of the Galactic pulsar population and of the PWN acceleration. 
Recent works have  investigated through simulations of pulsar populations the interpretation of AMS-02 data \cite{Cholis_2018,Manconi:2020ipm,Evoli_2021}. 
Here, we build upon previous works and extend them significantly in various novel aspects. We simulate a large number of realizations for Galactic pulsar populations, implementing different updated models for the source distribution, particle injection and propagation which reproduce ATNF catalog observations, instead of ad-hoc realizations of pulsar properties. For each mock galaxy, we compute the CR $e^+$  flux at the Earth from the resulting PWNe population. We then fit our predictions to the  AMS-02 data in order to determine the physical parameters of these populations. We also inspect the properties of individual sources which are able to explain the observed  $e^+$ flux. 
We simulate sources following different spatial distributions in the Galactic spiral arms, and investigating the effect of the radial distribution of sources. 
We also asses the impact of different assumptions on the spin-down properties, of propagation scenarios and  $e^+$ emission properties. Finally, we measure the average number of sources necessary to fill the gap between the  $e^+$ secondary flux and AMS-02 data. 

The $e^{\pm}$ fluxes from Galactic
pulsars are expected to contribute to the data on the $e^+ + e^-$ spectrum as well. 
Many authors in recent years (e.g. \cite{2019_Manconi,Evoli_2021}) included the contribution from catalog pulsar or simulations to model the $e^+ + e^-$ data of Fermi-LAT \cite{Abdollahi:2017nat}, AMS-02 \cite{PhysRevLett.122.101101,PhysRevLett.122.041102}, DAMPE \cite{2017_DAMPE} and  CALET \cite{PhysRevLett.120.261102}.  
However, the AMS-02 $e^+$ data up to 1~TeV are currently the most precise observable to constrain the characteristics of Galactic pulsar populations, being this the main scope of the present paper. 
The $e^+ + e^-$ data is indeed dominated by the $e^-$ produced by supernova remnants (SNR), that should be carefully modeled. This is beyond the scope of this paper, also because of the increasing of the number of variables that should  be taken into account.
The remainder of the paper is organized as follows. In Section~\ref{sec:positrons}
 our modeling for the $e^\pm$ production from a single PWN, and the basics of $e^\pm$ propagation in our Galaxy are illustrated. 
 The strategies to simulate the  Galactic pulsar populations and the different setups investigated are introduced in Section~\ref{sec:simulations}. The fit to AMS-02 data and the consequences for the pulsar characteristics of our minimization analysis are illustrated in Section~\ref{Results}, before concluding in Section~\ref{sec:closing}.

\section{Positrons from Galactic pulsars}
\label{sec:positrons}

In this Section we illustrate our model for the $e^\pm$ production from PWNe, 
while the strategies to simulate a Galactic pulsar population will be introduced in the next Section. 
We also briefly remind here the basics of $e^\pm$ propagation in our Galaxy. 

\subsection{Injection of $e^\pm$ from pulsars}
\label{sec:injection}
We recall here the main aspects for the emission of $e^{\pm}$  from pulsars and their PWNe into the ISM. The formalism is presented in ~\cite{DiMauro:2019yvh}, to which we refer for further details. 
Pulsars are rotating neutron stars with a strong surface magnetic field, and magnetic dipole radiation is believed to provide a good description for its observed loss of rotational energy \cite{Bykov:2017xpo,Amato:2020zfv}. We consider a model in which $e^{\pm}$ are continuously injected at a rate that follows the pulsar spin-down energy. \footnote{While 'pulsar' and 'PWN' have both been used in  literature to indicate the $e^\pm$ source, we here use 'pulsar' when referring to the simulated properties of pulsar at birth, while PWN is used when referring to the $e^\pm$ primary source.}
This scenario is indeed required to generate the TeV photons detected by Milagro and HAWC for Geminga and Monogem \cite{Abeysekara:2017hyn, DiMauro:2019yvh, Yuksel:2008rf}. 
The injection spectrum $Q(E,t)$ of $e^\pm$ at energy $E$ and time $t$ is described as:
\begin{equation}
\label{eq:spectrum}
    Q(E,t)=L(t) \left(\frac{E}{E_0} \right)^{-\gamma_e} \exp \left(- \frac{E}{E_c} \right) 
\end{equation}
where the cut-off energy $E_c$ is fixed at $10^5$ TeV, $E_0 = 1$  GeV and $\gamma_e$ is the $e^\pm$ spectral index. The magnetic dipole braking $L(t)$ is described by the function:
\begin{equation}\label{eq:L}
L(t)=\frac{L_0}{\left( 1+ \frac{t}{\tau_0}\right)^{\frac{n+1}{n-1}}}
\end{equation}
where $\tau_0$ is the characteristic time scale and $n$ defines the magnetic braking index.
Alternatively the injection spectrum can be parametrized with a broken power-law \cite{Abeysekara:2017science,DiMauro:2019yvh}, with a break at energies of the order of tens to hundreds GeV, a slope below the break $\approx 1.4$ and above the break $\approx 2.2$,  which  is compatible with multiwavelenght observations of PWNe, but with large uncertainties on the  parameters \cite{Torres:2014iua}. Here we adopt an effective approach using eq.~\ref{eq:spectrum}, avoiding the increasing of the number of degrees of freedom. The possible effects of this assumption will be discussed in Section~\ref{sec:characheristics}.
The total energy emitted by the source only into $e^+$ is given by:
\begin{equation}\label{eq:Etot}
    E_{tot}=\eta W_0= \int_{0}^T dt \int_{E_1}^{\infty} dE E Q(E,t)
\end{equation}
through which we obtain the value of $L_0$, fixing $E_1$=0.1 GeV \cite{Sushch:2013tna, Buesching:2008hr}. The parameter $\eta$ encodes the efficiency of conversion of the spin-down energy into $e^+$(which is half of the efficiency of conversion into $e^\pm$).
$W_0$ is the initial rotational energy of a pulsar with a moment of inertia $I$ (typically assumed to be $10^{45}$g cm$^2$, as obtained from canonical neutron star values) and rotational frequency $\Omega_0=2\pi/P_0$:
\begin{equation}\label{eq:Erot}
    W_0=E_{\rm rot,0}=\frac{1}{2} I {\Omega_{0}}^2\,.
\end{equation}
The spin-down luminosity $\dot{E} = dE_{\rm rot}/dt$ of a pulsar is the rate at which the rotational kinetic energy is dissipated:
\begin{equation}\label{eq:dotEP}
    \dot{E}=\frac{d E_{\rm rot}}{dt}= I \Omega \dot{\Omega}= -4\pi^2 I \frac{\dot{P}}{P^3}\,. 
\end{equation}
Assuming a small deviation from the dipole nature of the magnetic field  $B$ of the pulsar, the evolution of the star may be parameterized as  \cite{Ridley_2010}:
\begin{equation}\label{eq:PdotPB}
P^{n-2}\dot{P} = a k (B \sin \alpha)^2\,.
\end{equation}
where the angle $\alpha>0$ describes the inclination of the magnetic dipole with respect to the rotation axis, $a$ is a constant of unit s$^{n-3}$ and $k$ takes the value of $9.76 \times 10^{-40}$ s\,G$^{-2}$ for canonical characteristics of neutron stars. 
The spin-down luminosity evolves with time $t$ as in eq.~\ref{eq:L}:
\begin{equation}\label{eq:dotEevol}
    \dot{E}(t) = \dot{E}_0 \left( 1+ \frac{t}{\tau_0}\right)^{-\frac{n+1}{n-1}}.
\end{equation}
From this equation, one can notice that the pulsar has roughly a constant energy output from its birth till 
$t=\tau_0$, when the energy output starts to decrease as $\dot{E}\sim t^{-(n+1)/(n-1)}$. Finally, 
the prediction on $\tau_0$ is derived to be:
\begin{equation}\label{eq:tau0}
    \tau_0= \frac{P_0}{(n-1)\dot{P}_0}.
\end{equation}

In our benchmark model we will consider only sources with ages above 20 kyr, since $e^\pm$ accelerated to TeV energies in the termination shock are believed to be confined in the nebula or in the SNR until the merge of this system with the ISM, estimated to occur some kyr after the pulsar formation \cite{2011ASSP...21..624B}. We thus leave out sources for which the $e^\pm$  pairs might be still confined in the parent remnant. However, this effective treatment does not account for possible spectral or time-dependent modifications of the released particles. To understand the consequences of this assumption on the interpretation of the AMS-02 $e^+$ flux, we also test the hypothesis that only the $e^\pm$ produced after the escaping of the pulsar from the SNR contribute to the flux at the Earth. Following \cite{Evoli_2021}, we define $t_{BS}$ as the time at which the source leaves the parent SNR due to its proper motion and eventually forms a bow-shock nebula. The escape time of the pulsar from the remnant is described by: 
\begin{equation}
    t_{BS} \simeq 56 \left(\frac{E_{SN}}{10^{51} \text{erg}}\right)^{\frac{1}{3}} \left( \frac{n_0}{3 \text{ cm}^{-3}}\right)^{-\frac{1}{3}} \left( \frac{v_k}{280 \text{ km/s}} \right)^{-\frac{5}{3}} \text{kyr}
    \label{eq:bow_shock}
\end{equation}
where $n_0$ is the ISM density taken to be 3 or 1 cm$^{−3}$, $E_{SN}$ = $10^{51}$ erg is the energy emitted by the SN explosion and $v_k$ is the birth velocity of the pulsar. The formalism is reported in ~\cite{van_der_Swaluw_2003}, to which we refer for further details.

\subsection{Propagation of $e^\pm$ to the Earth}
\label{sec:propagation}
Once charged particles are injected in the Galaxy, they can propagate and eventually reach the Earth. 
We remind here very briefly the way we treat Galactic propagation of $e^\pm$, and address the reader to ~\cite{2010A&A...524A..51D,Manconi:2018azw} for further details. 
The  number density per unit time, volume $N_e(E, {\bf r}, t)$ of $e^{\pm}$ at an observed energy $E$, a position ${\bf r}$ in the Galaxy, and time $t$, which is the solution to the propagation equation considering only diffusion and energy losses, is given by \cite{DiMauro:2019yvh}:
\begin{equation}\label{eq:numberdensity}
    N_e(E, {\bf r}, t)=\int_0^t dt' \frac{b(E_s)}{b(E)} \frac{1}{(\pi \lambda^2(t',t,E))^{\frac{3}{2}}}  \exp \left(-\frac{| {\bf r}-{\bf r_s}|^2}{\lambda ^2 (t',t,E)} \right) Q(E_s,t')
\end{equation}
where the integration over $t'$ accounts for the PWN releasing $e^{\pm}$ continuously in time. The energy $E_s$ is the initial energy of $e^{\pm}$ that cool down to $E$ in a loss time $\Delta \tau$:
\begin{equation}\label{eq:losstime}
    \Delta \tau \equiv \int_E^{E_s} \frac{dE'}{b(E')} = t - t_{obs}.
\end{equation}
The $b(E)$ term is the energy loss rate, ${\bf r_s}$ indicates
the source position, and $\lambda$ is the typical propagation length defined as:
\begin{equation}\label{eq:lambda}
    \lambda^2 = \lambda^2(E,E_s) \equiv 4\int_E^{E_s} dE' \frac{D(E')}{b(E')} 
\end{equation}
where $D(E)=D_0 E^{\delta}$ is the diffusion coefficient taken as a power-law in energy. The $e^{\pm}$ energy losses include ICS off the ISRF and the synchrotron emission on the Galactic magnetic field.
The flux of $e^\pm$ at the Earth for a source of age $T$ and distance $d=|{\bf r_{\odot} - r_s}|$ is given by:
\begin{equation}
 \Phi_{e^\pm}(E) = \frac{c}{4\pi} \mathcal{N}_e(E,{\bf r=r_{\odot}},t=T).
 \label{eq:flux}
\end{equation}
In our analysis we will consider as benchmark case the propagation parameters as derived in \cite{dimauro2021multimessenger} from a fit to the latest AMS-02 data for the B/C, antiproton and proton data. We will label this model as {\it Benchmark-prop}, where 
 $D_0=0.042$ kpc$^2$/Myr and $\delta=0.459$. The value of $L$ is fixed to $4$ kpc, which is compatible with the recent results of ~\cite{Weinrich:2020ftb}.
Energy losses are computed on the interstellar photon populations at different wavelengths following \cite{Vernetto:2016alq}, by taking into account the Klein-Nishina formula for ICS, and on the Galactic magnetic field with intensity $B=3$ $\mu$G. 
As a comparison, we will also implement the {\it SLIM-MED} model derived in 
\cite{genolini2021new}, with the ISRFs taken from \cite{2010A&A...524A..51D} and $B=1$~$\mu$G. This model assumes $D_0=0.036$ kpc$^2$/Myr, $\delta=0.499$ and $L=4.67$ kpc.
However, the  parameter $L$ is not relevant in this study since we implement solutions without boundaries both in the radial and the vertical directions. The infinite halo approximation has been widely used to compute the flux from single sources located in the Galactic plane in ~\cite{Fornieri_2020,Recchia_2019,Manconi:2020ipm}.

\section{Simulations of Galactic pulsar populations}
\label{sec:simulations}

\begin{table*}[t]
\centering
\begin{tabular}{ c @{\hspace{10px}} | c @{\hspace{10px}} | c @{\hspace{10px}}  |c @{\hspace{10px}}  } \hline\hline
\textbf{Pulsar} 
&   \textbf{Simulated }                        &  \textbf{ Benchmark  }  
&    \textbf{ Variations }  \\ 
\textbf{property} 
&   \textbf{quantity }                        &    
&    \textbf{ }  \\ 
\hline
Age & $T$ &  Uniform  [0, $t_{max}$]  & - \\ 
\hline 
 & &  \texttt{CB20}\cite{Chakraborty:2020lbu} &   \texttt{FK06}\cite{2006ApJ...643..332F} \\
& $P_0$  &  Gaussian [0.3s; 0.15s] & - \\
Spin-down &$\log_{10}(B)$  & Gaussian [12.85G; 0.55G]   & Gaussian [12.65G; 0.55G]\\
&  $n$ & Uniform [2.5-3] & Constant [3]  \\
&  cos$\alpha$  & Uniform [0-1]& Constant [0]  \\
\hline 
$e^\pm$ injection & $\gamma_e$  & Uniform [1.4-2.2]   & - \\ 
 &  $\eta$ &  Uniform [0.01-0.1]  & - \\ 
\hline 
Radial & ${\bf r}$ & $\rho_L(r)$ \cite{Lorimer_2006}  &   $\rho_F(r)$\cite{2006ApJ...643..332F} \\
distribution &  &    & \\
\hline 
Kick velocity & $v_k$ & -   & \texttt{FK06VB} \cite{2006ApJ...643..332F}  \\  \hline\hline
\end{tabular}
\caption{Summary of the  parameters from which we build the mock pulsar catalogs. We report the pulsar simulated quantities (first two columns), the distributions adopted in their simulation, with the boundary of their validity range, for our benchmark case (\textbf{\texttt{[ModA]}}, third column), as well as the tested 
variations (last column).
See Section~\ref{sec:simulations} for details.}
\label{tab::models}
\end{table*}

We simulate Galactic pulsars following the injection and propagation model described in Section~\ref{sec:positrons}. 
For each realization we compute the $e^+$ flux from every PWN. In what follows, we describe the strategy adopted to produce the mock catalogs and the five simulation setups considered (\texttt{ModA-B-C-D-E}) here.
In each simulation, the total number of sources is fixed at $N_{\rm PSR} = t_{max} \dot{N}_{PSR}$, where $t_{max}$ is the maximum simulated age and $\dot{N}_{PSR}$ is the pulsar birth rate. Different estimates for the Galactic $\dot{N}_{PSR}$ range from one to four per century \cite{2004IAUS..218..105L,Keane:2008jj,2006ApJ...643..332F}. We here assume the maximum age of the sources to be $t_{max}$ = $10^8$ yr, and $\dot{N}_{PSR}$ = 0.01 yr$^{-1}$.  However, we have checked that $t_{max}$ = $10^9$ yr does not change the conclusions of this paper.

In order to compute the $e^+$ flux at the Earth for each mock source, we need to specify its position in the Galaxy, its age and the source term  $ Q(E,t)$ (see eq.~\ref{eq:spectrum}). Specifically, the fundamental parameters 
of each simulation are:  $T$,  $P_0$, $B$, $n$, $\alpha$, $\gamma_e$, $\eta$  and the position ${\bf r}$ in Galactocentric coordinates. For a specific set of simulations, we also simulate the birth-kick velocity 
$v_k$. A summary of the simulated quantities is illustrated in Table~\ref{tab::models} and outlined in what follows.

First of all, the simulation assigns to each mock pulsar an age $T$ extracted uniformly between $t=0$ and $t_{max}$. Then, by extending the functions implemented in the Python module {\tt gammapy.astro.population} \cite{Nigro_2019,CTAConsortium:2017xaq}, we sample the values of $P_0, B, n$ and $\alpha$ from the distributions provided in \cite{Chakraborty:2020lbu} (\texttt{CB20}), which will be our benchmark model.
Specifically, $P_0$ is simulated according to a Gaussian distribution with $P_{0,mean}$ = 0.3 s and $P_{0,std}$ = 0.15 s. We also impose a lower bound on $P_0$ = 0.83 ms, as physically motivated in \cite{LATTIMER_2007}.
The magnetic field is simulated following a Gaussian distribution for  $\log_{10}(B)$, with  $\log_{10}(B)_{mean}$ = 12.85 G and $\log_{10}(B)_{std}$ = 0.55~G. 
The values of $n$ and $\cos{\alpha}$ are taken from uniform distributions, respectively in the range [2.5-3] and [0-1] according to \cite{Chakraborty:2020lbu}. We note that from the simulated values of $P_0, B, n, \alpha$ we derive for each pulsar $W_0$ and $\tau_0$ through eqs.~\ref{eq:Erot} and \ref{eq:tau0}.
Since the spectral index $\gamma_e$ of accelerated particles is uncertain, and may vary significantly for each  PWN \cite{Gaensler:2006ua, Mlyshev:2009twa}, it is sampled from uniform distributions within [1.4-2.2]. 
Finally, the value of $\eta$ for each source is sampled from a uniform distribution in the range [0.01-0.1].

In order to assess the effect of different distributions for $P_0, B, n$ and $\alpha$, we  consider the model in \cite{2006ApJ...643..332F} (\texttt{FK06} hereafter).  
While $P_0$ follows the same distribution of \texttt{CB20}, $\log_{10}(B)$ is taken from a Gaussian distribution with $\log_{10}(B)_{0,mean}$ = 12.65 G and $\log_{10}(B)_{0,std}$ = 0.55~G, while $n$=3 and $\sin\alpha$=1 for each source. We note that both \texttt{CB20} and \texttt{FK06} models have been calibrated to reproduce the characteristics of the sources detected in the ATNF catalog \cite{2005AJ....129.1993M}, like the $P$, $\dot{P}$, $B$, flux densities at 1.4 GHz, Galactic longitudes and Galactic latitudes distributions. \texttt{CB20} is the most updated model and considers the variation of more parameters with respect to \texttt{FK06}. 

To test the scenario described by eq.~\ref{eq:bow_shock}, we additionally simulate for each source its birth-kick velocity, adopting its distribution as reported in \cite{2006ApJ...643..332F} (\texttt{FK06VB}) and implemented in  {\tt gammapy.astro.population} \cite{Nigro_2019,CTAConsortium:2017xaq}, which is the sum of two Gaussians (see their eq. 7) for each of the 3 velocity components. 

\subsection{Spatial distribution of pulsars in the Galaxy }
\label{sec:spatial_distribution}
To complete the construction of the mock catalogs of Galactic sources the position ${\bf r}$ of each pulsar has to be determined.
Using {\tt gammapy.astro.population} \cite{Nigro_2019,CTAConsortium:2017xaq} we adopt the radial surface density of pulsars $\rho_L(r)$ proposed by \cite{Lorimer_2006}:
\begin{equation}\label{eq:rho_L}
    \rho_L(r)=A_1\left(\frac{r}{r_{\odot}}\right) \exp \left[-C\left(\frac{r-r_{\odot}}{r_{\odot}}\right)\right]. 
\end{equation}
As a comparison, we also consider the radial surface density $\rho_F(r)$  in \cite{2006ApJ...643..332F}:
\begin{equation}\label{eq:rho_F}
    \rho_F(r)=A_2 \frac{1}{\sqrt{2 \pi}\sigma} \exp \left(-\frac{(r-r_{\odot})^2}{2 \sigma^2 }\right).
\end{equation}
See \cite{Lorimer_2006, 2006ApJ...643..332F} for the values of the parameters. We sample the position ${\bf r}$ of each source combining the radial surface density with the  spiral arm structure of the Milky Way of ~\cite{2006ApJ...643..332F} (see their Table 2 for the spiral arm parameters), 
as implemented in {\tt gammapy.astro.population} \cite{Nigro_2019,CTAConsortium:2017xaq}. We test only one spiral arm structure, since the most important aspect in the computation of the $e^+$ flux is the source density in the arms nearby the Sun, instead of the position of the arms themselves.
The distance of each source is $d$=$|{\bf r}-{\bf r}_{\odot}|$, with ${\bf r}_{\odot} = (8.5, 0, 0) {\rm kpc}$.

\begin{figure*}[t]
    \centering
    \begin{subfigure}[b]{0.5\textwidth}
        \centering
        \includegraphics[width=1\textwidth]{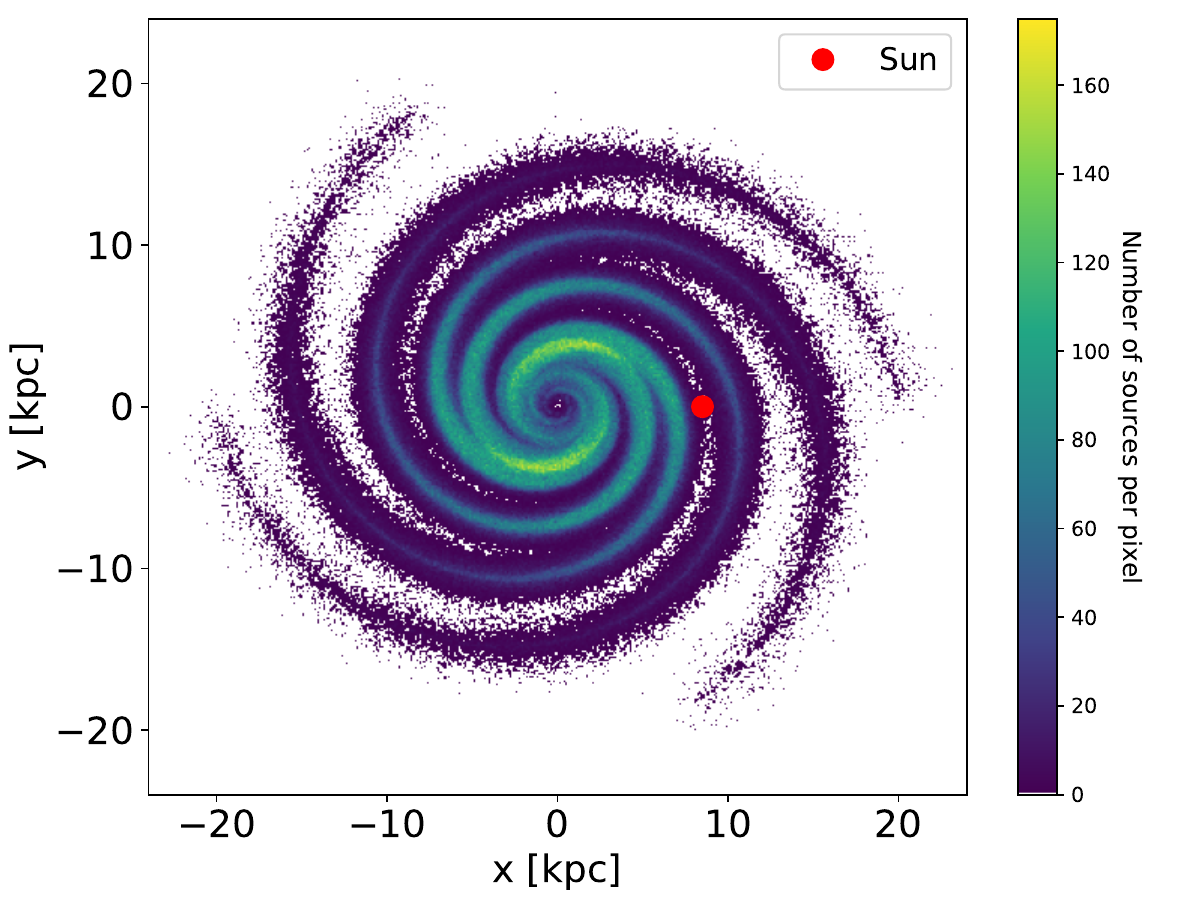}
        \caption{}
    \end{subfigure}%
    ~ 
    \begin{subfigure}[b]{0.5\textwidth}
        \centering
        \includegraphics[width=1\textwidth]{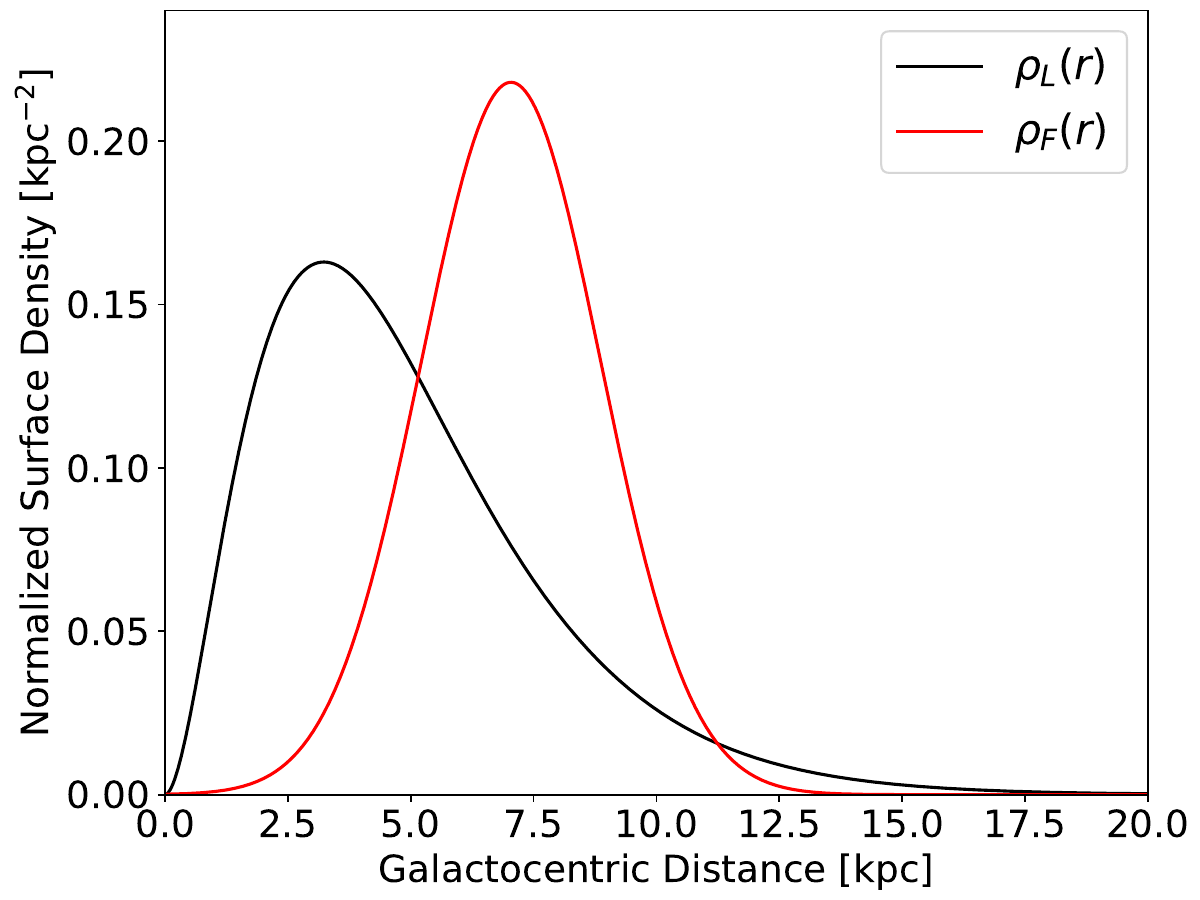}
        \caption{}
    \end{subfigure}
    \caption{Panel (a): Position of pulsars in the Galactic plane (grouped in pixels of size 0.015 kpc$^2$) for one realization of our Galaxy obtained with $\rho_L(r)$ radial surface density~\cite{Lorimer_2006} and the spiral-arm model of ~\cite{2006ApJ...643..332F}. The color bar indicates the number of sources in each pixel. In panel (b) are reported the normalized radial surface densities  $\rho_L(r)$ (\cite{Lorimer_2006},  black line) and $\rho_F(r)$ (\cite{2006ApJ...643..332F}, red line).}
    \label{Fig M1}
\end{figure*}

In Figure~\ref{Fig M1}(a) we report the positions in the Galactic plane of the mock sources, for one 
configuration of our Galaxy, adopting the $\rho_L(r)$ radial surface density. 
Due to the fast energy-losses that affects $e^\pm$,
the most relevant contribution to the $e^+$ flux will come from the two spiral arms that surround the Earth and that are named Sagittarius and Orion.
In Figure~\ref{Fig M1}(b) we also display the $\rho_L(r)$ and $\rho_F(r)$ profiles reported in eq.~\ref{eq:rho_L} and \ref{eq:rho_F} (normalized in order to have $\int_0^{+\infty}\rho_i(r)dr=1$ with $i=L,F$).
We note that  $\rho_L(r)$ is similar to other radial distributions used in literature \cite{Case_1998,Yusifov_2004}, and we consider it as a good benchmark. The  $\rho_F(r)$ profile
 effectively maximizes the effects of different radial profiles on the $e^+$, by setting higher pulsar densities in the two spiral arms surrounding the Earth. 

\subsection{Summary of simulation setups} 
\label{sec:simulation_setups}
We recap and label here the combinations of the different simulation setups described above and listed in Table~\ref{tab::models}. 
\paragraph{\texttt{ModA} (benchmark).} 
 Spin-down and pulsar evolution properties are taken from \texttt{CB20} \cite{Chakraborty:2020lbu}, while the radial surface density of sources is modelled with $\rho_L(r)$ (see eq.~\ref{eq:rho_L},~\cite{Lorimer_2006}). $\eta$ and $\gamma_e$ are extracted from uniform distributions, while the propagation in the Galaxy is taking into account with {\it Benchmark-prop} model following ~\cite{dimauro2021multimessenger}. 
\paragraph{\texttt{ModB} (radial distribution effect).} 
 Same as \texttt{ModA} but with the radial surface density of sources $\rho_F(r)$ instead of $\rho_L(r)$ (see eq.~\ref{eq:rho_F},~\cite{2006ApJ...643..332F}). 
 \paragraph{\texttt{ModC} (spin-down properties effect).} 
  Same as \texttt{ModA}, but spin-down properties are taken from \texttt{FK06} \cite{2006ApJ...643..332F}. 
 \paragraph{\texttt{ModD} (propagation effect).} 
  Same as \texttt{ModA} apart for propagation in the Galaxy, modelled as in ~\cite{genolini2021new} (their model {\it SLIM-MED}).
   \paragraph{\texttt{ModE} (kick velocity effect).} 
  Same as \texttt{ModA}, but considering only the $e^{\pm}$ emitted after the escaping of pulsars from the SNR. The birth kick velocities are sampled adopting the distribution \texttt{FK06VB} reported in ~\cite{2006ApJ...643..332F}.

\section{Results}\label{Results}

The aim of this paper is to study the characteristics of the Galactic  pulsar population that may contribute to the $e^+$ flux measured by the AMS-02 experiment  \cite{PhysRevLett.122.041102} by using simulations. 
We describe here how we performed the fit to the data, and which are the physics results of our minimization analysis. 
For each simulation setup  described in Section~\ref{sec:simulation_setups}, we build and test 1000 simulations. All the results reported in this Section refer to our benchmark \texttt{ModA}, if not differently stated.
 We compute the $e^+$ flux at the Earth as the sum of the primary component due to PWNe emission
 (see Section~\ref{sec:positrons} and  Section~\ref{sec:simulations}), and a secondary component due to the fragmentation of CRs on the nuclei of the ISM, taken  from \cite{dimauro2021multimessenger} or \cite{genolini2021new} consistently with the propagation model employed. 
The secondary component enters in our fits with a free normalization factor $A_S$, which we generously let to vary between 0.01 and 3. 
We also let the total flux generated by the sum of all PWNe to be shifted by an overall normalization factor $A_P$. The values of $A_P$ and $A_S$ are obtained for each simulation with the fit procedure.

We fit AMS-02 data \cite{PhysRevLett.122.041102} above 10 GeV, in order to avoid strong influence from solar modulation and other possible low energy effects \cite{Boudaud:2016jvj}. Nevertheless, we correct our predictions for solar modulation effects following the force field approximation and leaving the Fisk potential $\phi$ free to vary between 0.4 and 1.2 GV. 
The comparison of our predictions with the AMS-02 $e^+$ data is performed by a standard $\chi^2$ minimization procedure.
We neglect the presence of correlations in the systematic errors of AMS-02 data points since the Collaboration has not provided them \cite{PhysRevLett.122.041102}. Moreover, we do not think the smoothness of the AMS-02 data, that is the main characteristic of the observed flux that will guide our results, would be modified significantly by such correlations.

\subsection{Comparison to the AMS-02 $e^{+}$ data}
\label{sec:results_fit}

The fit of the predictions for the total $e^+$ flux to the AMS-02 data is performed for all the 1000 simulations built for each scenario  \texttt{A-B-C-D-E}.
In Table~\ref{tab:Table_1} we report the number of simulations, out of 1000, that produce different values of the reduced chi squared $\chi^2/d.o.f.=\chi^2_{\rm red}$ for each simulation setup.

\begin{table}[t]
\begin{center}
\begin{tabular}{ |c|c|c|c| } 
 \hline
         & $\chi^2_{\rm red}<2$ & $\chi^2_{\rm red}<1.5$ & $\chi^2_{\rm red}<1$ \\
 \hline
 \texttt{ModA} &        15       &          8         &        4   \\ 
 \texttt{ModB} &        30       &          19        &        6   \\ 
 \texttt{ModC} &        15       &          10        &        3   \\ 
 \texttt{ModD} &        42       &          25        &        10  \\ 
  \texttt{ModE} &        4       &          2        &        2  \\ 
 \hline
\end{tabular}
\caption{Number of simulations (out of 1000) that produce a  $\chi^2_{\rm red}$ smaller than 2, 1.5 or 1 in the fit to AMS-02 data \cite{PhysRevLett.122.041102}, for each simulation setup.}
\label{tab:Table_1}
\end{center}
\end{table}

The difference between \texttt{ModA} and \texttt{ModD} is relative only to the propagation and the energy losses models. With \texttt{ModD} we obtain a higher number of simulations compatible with the data: the {\it SLIM-MED} model produces indeed fluxes from single sources which are smoother with respect to {\it Benchmark-prop}, and a little bit higher at lower energies. 
They better accommodate the AMS-02 spectrum. In all the tested setups, the number of mock galaxies with a $\chi^2_{\rm red}<1$ (2) does not exceed 1\% (4\%).

\begin{figure*}[h]
    \centering
    \begin{subfigure}[b]{0.5\textwidth}
        \centering
        \includegraphics[width=1\textwidth]{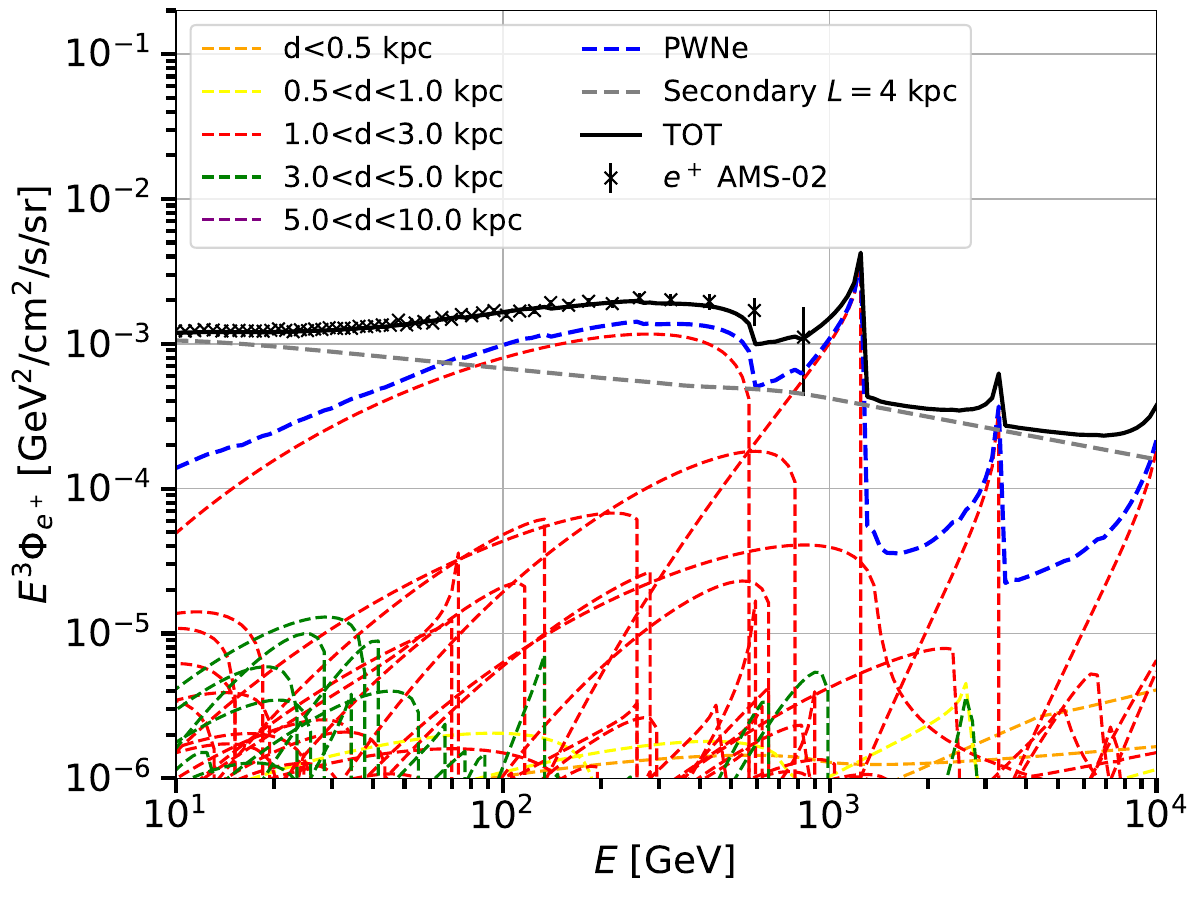}
        \caption{}
    \end{subfigure}%
    ~ 
    \begin{subfigure}[b]{0.5\textwidth}
        \centering
        \includegraphics[width=1\textwidth]{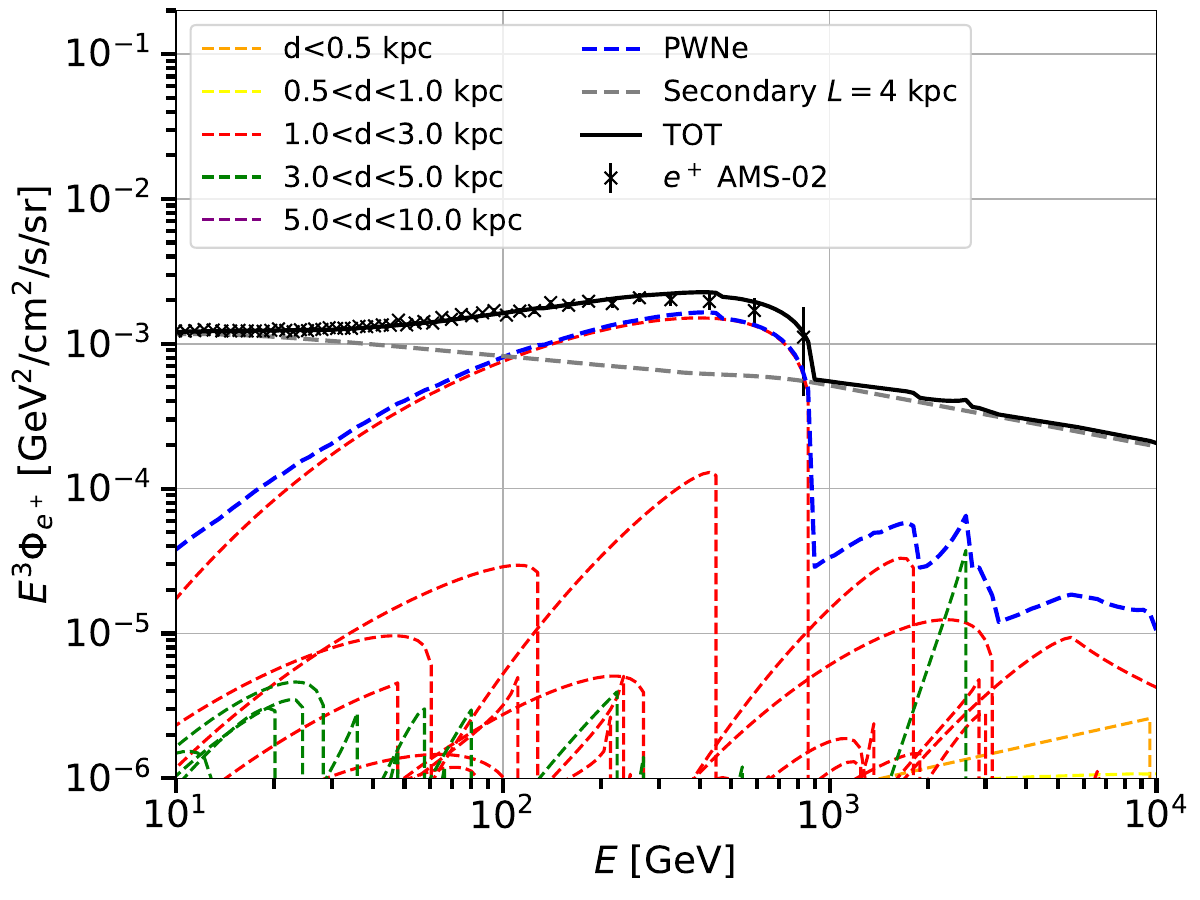}
        \caption{}
    \end{subfigure}
    \caption{Comparison between the AMS-02 $e^+$ flux data \cite{PhysRevLett.122.041102} (black points) and the  flux  from secondary production (grey dashed line) and PWNe (blue dashed line) for two \texttt{ModA} realizations of the Galaxy with $\chi^2_{\rm red}<1$. The contributions from each source, reported with different colors depending on their distance from the Earth, are shown.}
    \label{Fig R1}
\end{figure*}

In Figure~\ref{Fig R1} we plot the $e^+$ flux obtained for two illustrative simulated galaxies with $\chi^2_{\rm red}<1$, within  \texttt{ModA}. The contributions from each PWN, from the secondary emission and their sum are shown along with the AMS-02 data. 
The contribution from PWNe is significant for energies above 10-20 GeV and dominant over 100 GeV and may have different features, in particular at unconstrained energies above 1 TeV, depending on the specific simulation. As we will discuss later, the number of sources that contribute to the observed spectrum is limited, from a few to O(10). 
We notice that the secondary flux, while decreasing with energy, practically forbids the realization of sharp cut-offs in the $e^+$ spectrum above TeV energies. 
The different features of the flux from single PWNe are due to the peculiar combination of the input parameters. In particular, the peaked shape is typically associated to small $\gamma_e$ and $\tau_0$ values. 

All the good fits to the data provide a value for $A_S$ between 2 and 2.5, which might be at least partially
ascribable to an underestimation of spallation cross sections \cite{Delahaye:2008ua}. 
As for the allowed overall normalization $A_P$ of the PWN primary flux, we find on average values slightly smaller than one.

\begin{figure*}[h]
    \centering
    \begin{subfigure}[b]{0.5\textwidth}
        \centering
        \includegraphics[width=1\textwidth]{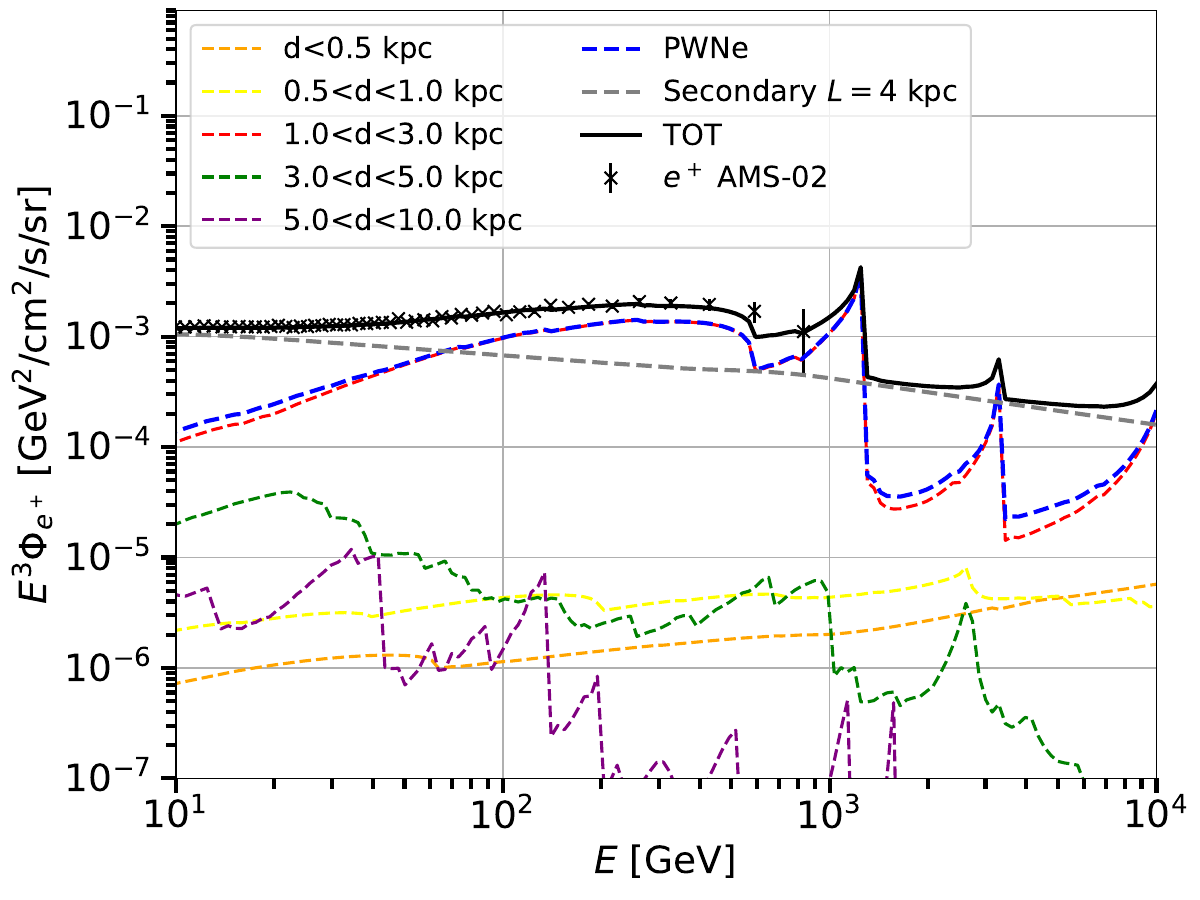}
        \caption{}
    \end{subfigure}%
    ~ 
    \begin{subfigure}[b]{0.5\textwidth}
        \centering
        \includegraphics[width=1\textwidth]{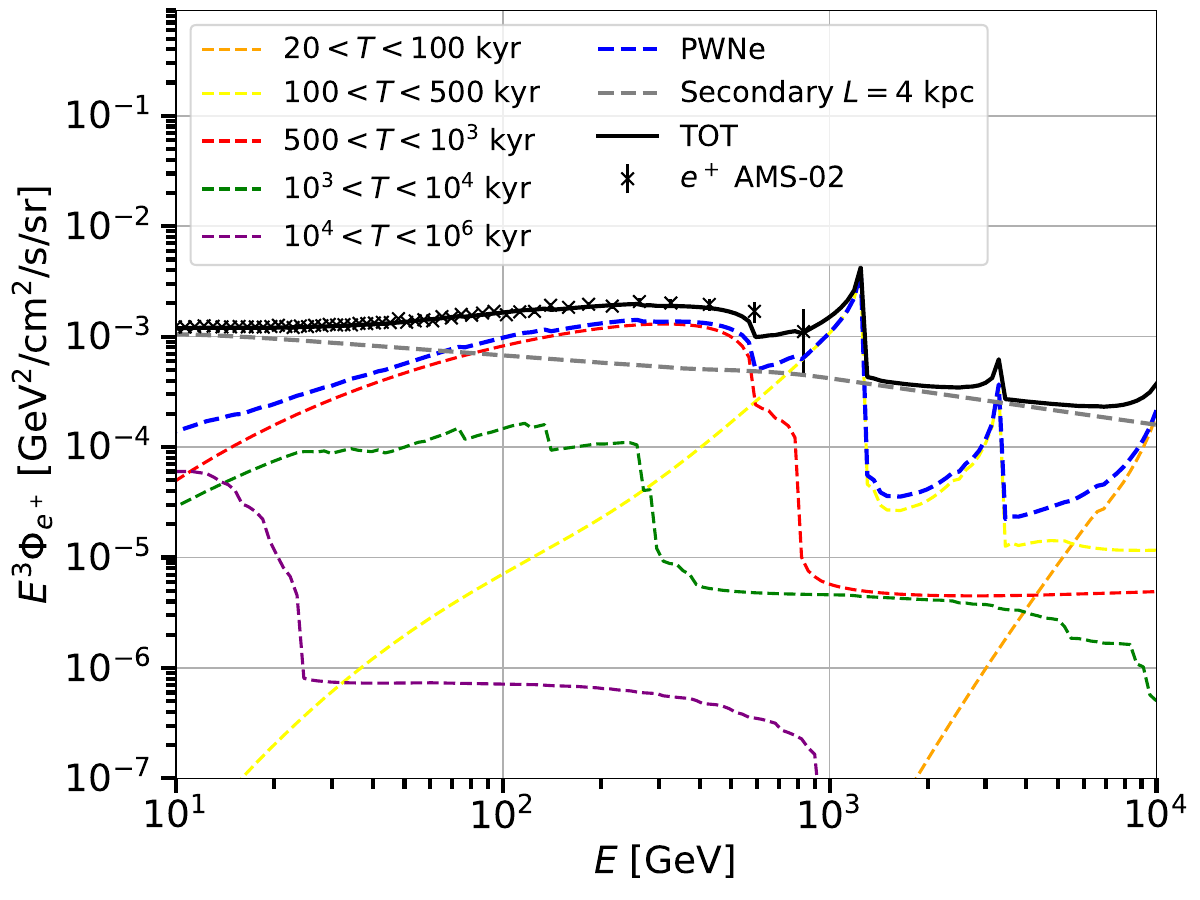}
        \caption{}
    \end{subfigure}
    \caption{Effect of distance and age of pulsars in a specific mock galaxy within setup \texttt{ModA}. 
    Panel a (b) reports the contribution to the $e^+$ flux for different distance (age) subsets. The dashed gray line reports the secondary flux, while the solid line corresponds to the total flux. AMS-02 data are from ref.~\cite{PhysRevLett.122.041102} (black points).}
    \label{Fig R2}
\end{figure*}

In order to understand the properties of the pulsar populations which fit the observations, we report 
in Figure~\ref{Fig R2} the contribution to the $e^+$ flux coming from pulsars grouped in different subsets of distance from the Earth (left) and age (right). In this realization,
the dominant contribution comes from the ring between 1 and 3 kpc. This result is the interplay between the presence of a spiral arm (see Figure~\ref{Fig M1}), which enhances the number of sources,  and  the typical propagation length of high-energetic $e^+$, affected by severe energy losses.
Despite the smaller effect from radiative cooling, the flux from sources within 1 kpc is lower due to the paucity of sources.

The division in age rings shows the scaling of the maximum energy $E_{\rm max}$ with the age of sources. In the Thomson approximation energy losses would provide $E_{\rm max}\propto 1/t$, inferred from $dE/dt \propto - E^2$. 
However, in the Klein-Nishina regime we observe a more complex behaviour. 
Pulsars older than $10^6$ kyr do not contribute significantly to the $e^+$ flux above 10 GeV, while the highest contribution around TeV energies come from sources younger than 500 kyr. 
We have checked that sources younger than 20 kyr do not produce sizeable effects on our analysis, and the energy range in which they would produce a relevant flux is well above AMS-02 data.

\begin{figure}[t]
\centering {
\includegraphics[width=0.60\textwidth]{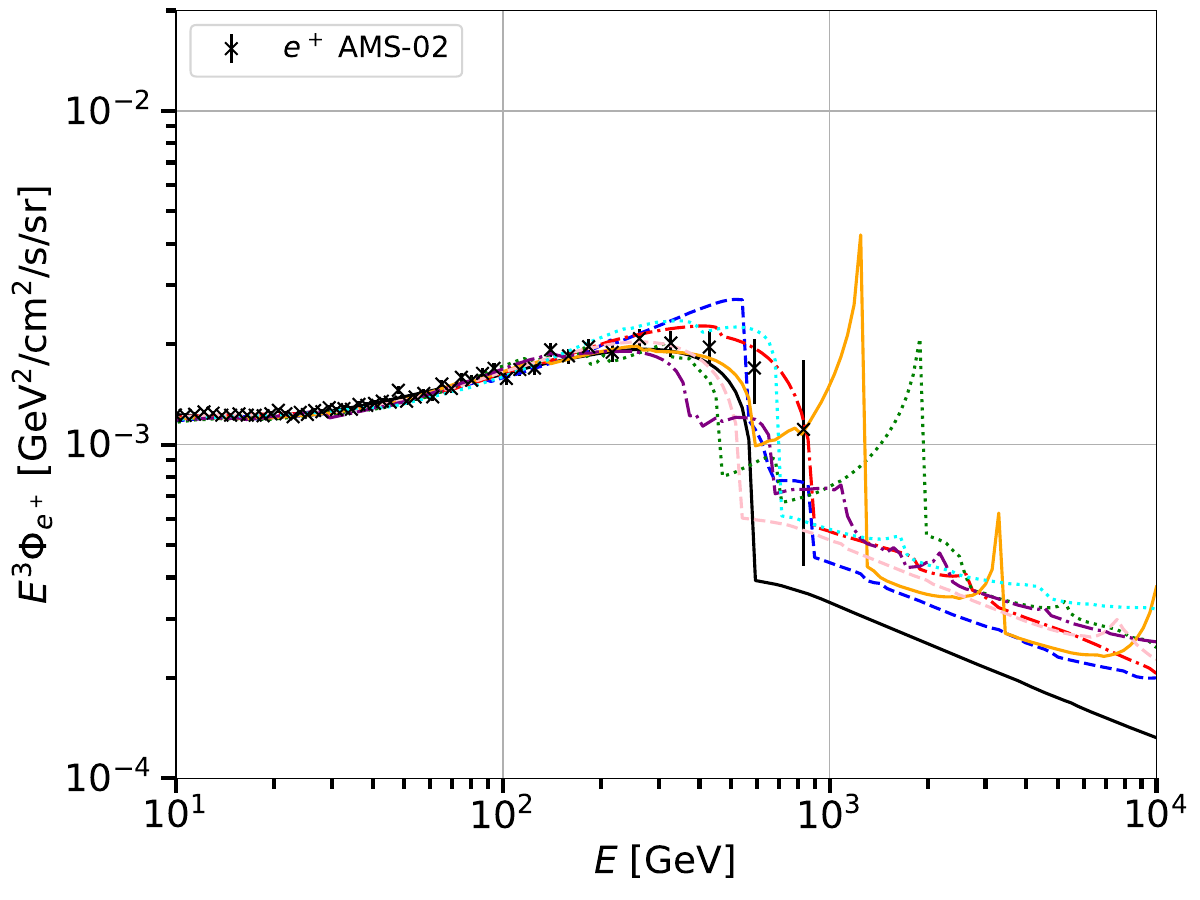}
}
\caption{Total (secondary plus PWNe) $e^+$ flux obtained from all the 8 simulations within \texttt{ModA} with $\chi^2_{\rm red}<1.5$, along with  AMS-02 data \cite{PhysRevLett.122.041102} (black points).}
\label{Fig R3}
\end{figure}

In order to inspect the effects of different simulated Galactic populations, we plot in Figure~\ref{Fig R3}  the total $e^+$ flux for all the pulsar realizations within \texttt{ModA}, and having $\chi^2_{\rm red}$<1.5 on AMS-02 data. 
For energies lower than 200 GeV, differences among the realizations are indistinguishable. The data in this energy range are very constraining. Instead, above around 300 GeV the peculiarities of each galaxy show up, thanks to the larger relative errors in the data.
Above 1 TeV the predictions are unconstrained by data. Nevertheless, all the simulations predict globally 
decreasing fluxes, as expected by energy losses and  continuous $e^\pm$ injection.
However, at these energies  the total flux gets dominated by the secondary component. 

Concerning the other simulation setups, we do not find significant differences between \texttt{ModA} and \texttt{ModC}, namely between \texttt{CB20} and \texttt{FK06} pulsar evolution models. For both cases on average the dominant contribution from PWNe comes from the 1-3 kpc distance ring. On the other hand, \texttt{ModB} gives a higher number of simulations that are compatible with the data. As shown in Figure~\ref{Fig M1}(b), the $\rho_F(r)$ radial distribution predicts a higher number of sources 
with respect to $\rho_L(r)$ in the spiral arms close to the Earth. Therefore, within \texttt{ModB} there is a higher probability to simulate sources close to the Earth with characteristics compatible with the AMS-02 data. However, the number of simulations 
with $\chi^2_{red}<1.5$ is <2\% for both \texttt{ModA} and \texttt{ModB}.

\texttt{ModE} differs from all the other cases due to the different computation of the fraction of $e^+$ produced by sources that actually contribute to the $e^+$ data (see Section~\ref{sec:injection}). We consider the ISM density $n_0$ = 3 cm$^{−3}$ in the computation of $t_{BS}$, see eq.~\ref{eq:bow_shock}. 
We find that, by fixing the maximum PWN efficiency of conversion of $W_0$ in $e^+$ to $50\%$, only 4 simulations fit the data with $\chi^2_{red}<2$ and all of them are dominated by a single powerful source, with $A_P$ $\sim$ 5, at the edge of the prior. We note that the renormalization factor $A_P$ is related to $\eta$, so that the actual efficiency of the single source $i$ is $A_P\times \eta_i$. 
Since this setup considers only the $e^+$ emitted after the escape of the pulsar from the SNR, the fit to the data selects the galaxies with sources that still have a great rotational energy that can be converted into $e^+$ at the exit time. 
Instead, if we do not put any upper limit for the parameter $A_P$, i.e.~to the efficiency, the number of simulations with $\chi^2_{red}<2$ increases to about one hundred. These unphysical values for the PWN efficiency  could be partially reduced by increasing the pulsar birth rate and so the number of pulsars in the simulations, or considering a different distribution of pulsar properties at birth which systematically predicts more energetic sources.

\subsection{Mean number of PWNe dominating the \textbf{$e^+$}  flux}
\label{sec:dominant_pulsar}

We inspect in this Section the average number of sources which contribute the most to the $e^+$ and thus can shape the AMS-02 flux. We adopt two complementary criteria  to estimate the number of sources that are responsible for the most significant contribution of the PWNe $e^+$ emission:
\begin{enumerate}
    \item {\em AMS-02 errors}: we count all the sources that produce a 
    flux higher than the experimental flux error in at least one energy bin above 10 GeV.
    \item {\em Total flux 1\%}: we count the sources that produce the integral of $\Phi_{e^\pm}(E)$ 
     between 10 and 1000 GeV higher than 1\% of the total integrated $e^+$ flux measured by AMS-02.  
\end{enumerate}

\begin{figure}[t]
 {
\includegraphics[width=0.5\textwidth]{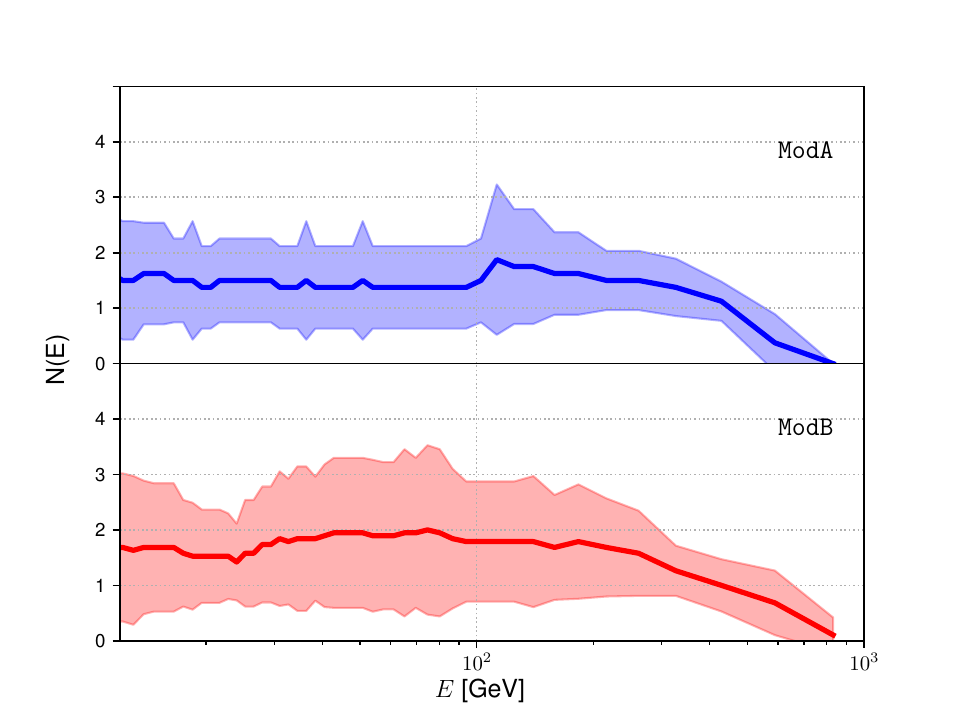}
\includegraphics[width=0.5\textwidth]{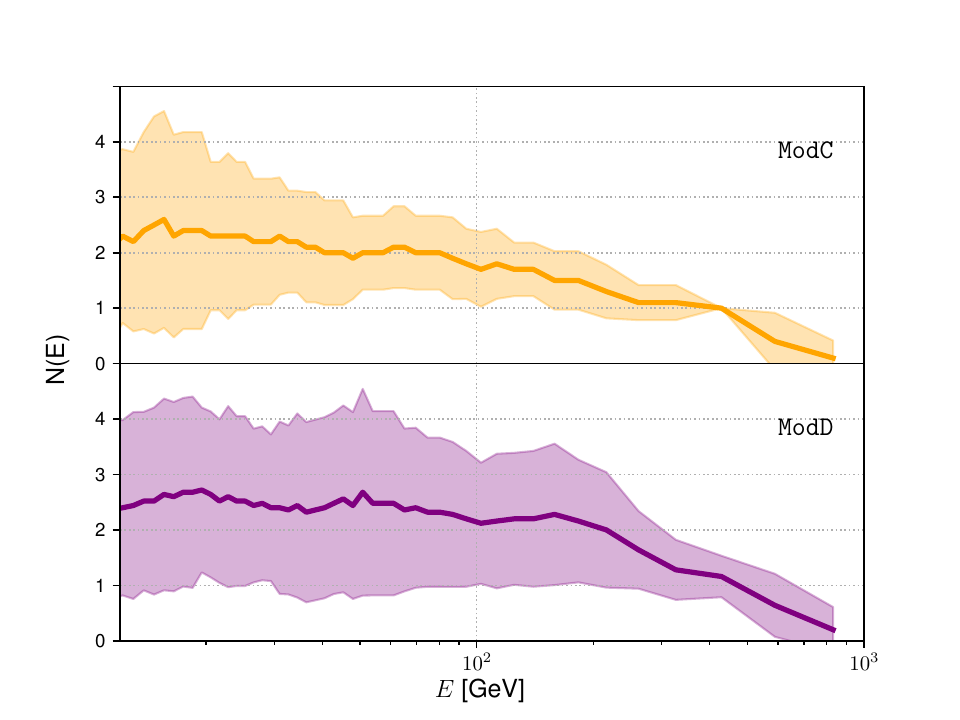}
}
\caption{Mean number of PWNe that satisfy the {\em AMS-02 errors} criterion in the single energy bin of AMS-02 data \cite{PhysRevLett.122.041102}. We also show the $68\%$ containment band for simulations with $\chi^2_{\rm red}<1.5$ (see the main text for further details).}
\label{Fig R4}
\end{figure}

In Figure~\ref{Fig R4} we report the average number of PWNe with the standard deviation (68\% containment band) that contribute in the different energy bins of AMS-02, for configurations with $\chi^2_{\rm red}<1.5$, adopting the {\em AMS-02 errors} criterion.
On average, 2-3 sources shine with a flux at least at the level of AMS-02 $e^+$ data errors.
We also find a decreasing number of dominant sources with increasing energy for all the setup reported. This result is partially induced by the larger experimental errors at high energy, which raise the threshold for the minimum flux that a PWN has to produce in order to satisfy the {\em AMS-02 errors} criterion. 
Moreover, being the age simulated in a uniform interval, the number of young sources responsible for the highest energy fluxes is smaller than for old pulsars, whose $e^+$ have suffered greater radiative cooling. 
Overall, it indicates that only a few sources with a large flux are required in order to produce a good fit to the data.

\begin{table}[t]
\begin{center}
\begin{tabular}{ |c|c|c| } 
 \hline
         &  AMS-02 errors   &  Total flux 1\%  \\ 
 \hline
 \texttt{ModA} &        1.3/2.9/3.3       &     1.0/1.8/2.2        \\ 
 \texttt{ModB} &        3.5        &     1.9        \\ 
 \texttt{ModC} &        3.9         &     3.0        \\  
 \texttt{ModD} &        5.4        &     3.5        \\ 
 \texttt{ModE} &        1.0        &     1.0        \\ 
 \hline
\end{tabular}
\caption{Average numbers of sources that satisfy the {\em AMS-02 errors} and  {\em Total flux 1\%} criteria, for all the galaxies within
each simulation setup, with $\chi^2_{\rm red}<1.5$. For \texttt{ModA}, results are provided also for $\chi^2_{\rm red}<1$ (left) and $\chi^2_{\rm red}<2$ (right).}
\label{tab:Table_2}
\end{center}
\end{table}

In Table~\ref{tab:Table_2} we report the average number of sources that satisfy the criteria listed above, for all the simulated galaxies which provide a good fit to AMS-02 data ($\chi^2_{\rm red}<1.5$). 
We obtain small numbers of sources responsible for most of the measured $e^+$, typically around 3, irrespective of the simulation scheme. 
Scenarios with a large number of sources explaining the CR $e^+$ data are disfavored. 
This result is due to the fact that AMS-02 measures a smooth flux, therefore several PWNe contributing at different energies would create wiggles in the total flux which are not detected. 
Instead, a few sources generating a flux that covers a wide range of energies produce a smooth contribution compatible with the data.

The values found around $2-2.5$ for the parameter $A_S$ can be justified by the preference of the data to have a high contribution of the featureless flux of secondary.
The change of propagation setup from \texttt{ModA} to \texttt{ModD}, produces an higher number of simulations that are compatible with the data, given the flux smoothing due to the alternative propagation setup. 
A slightly greater number of sources with respect to \texttt{ModA} satisfies the criteria for \texttt{ModD}. 
Dissecting results within \texttt{ModA}, we find that the mean number of sources decreases with decreasing $\chi^2_{\rm red}$, consistently with the requirement of a smooth trend of $e^+$ flux. 
In all the other simulation setups, except for \texttt{ModE}, we confirm the same trend. 
The \texttt{ModE} results have already been discussed in Section~\ref{sec:results_fit}. Summarizing, the two selection criteria {\em AMS-02 errors} and {\em Total flux 1\%}, 
whereas based on different quantitative assumptions, 
provide in practice very similar results.
This result is  complementary also to earlier analysis \cite{2010_Kawanaka} adopting a different strategy and working on PAMELA, FERMI-LAT and ATIC/PPB-BETS \cite{article_Chang,torii2008highenergy} data. There, a single or a few energetic pulsars are required to explain the measurements, and scenarios composed by multiple pulsars are found to be disfavored based on age and distance criteria, and by the data themselves.

\subsection{Characteristics of PWNe dominating the $e^+$ flux}\label{sec:characheristics}

\begin{figure*}[t]
    \centering
    \begin{subfigure}[b]{0.5\textwidth}
        \centering
        \includegraphics[width=1\textwidth]{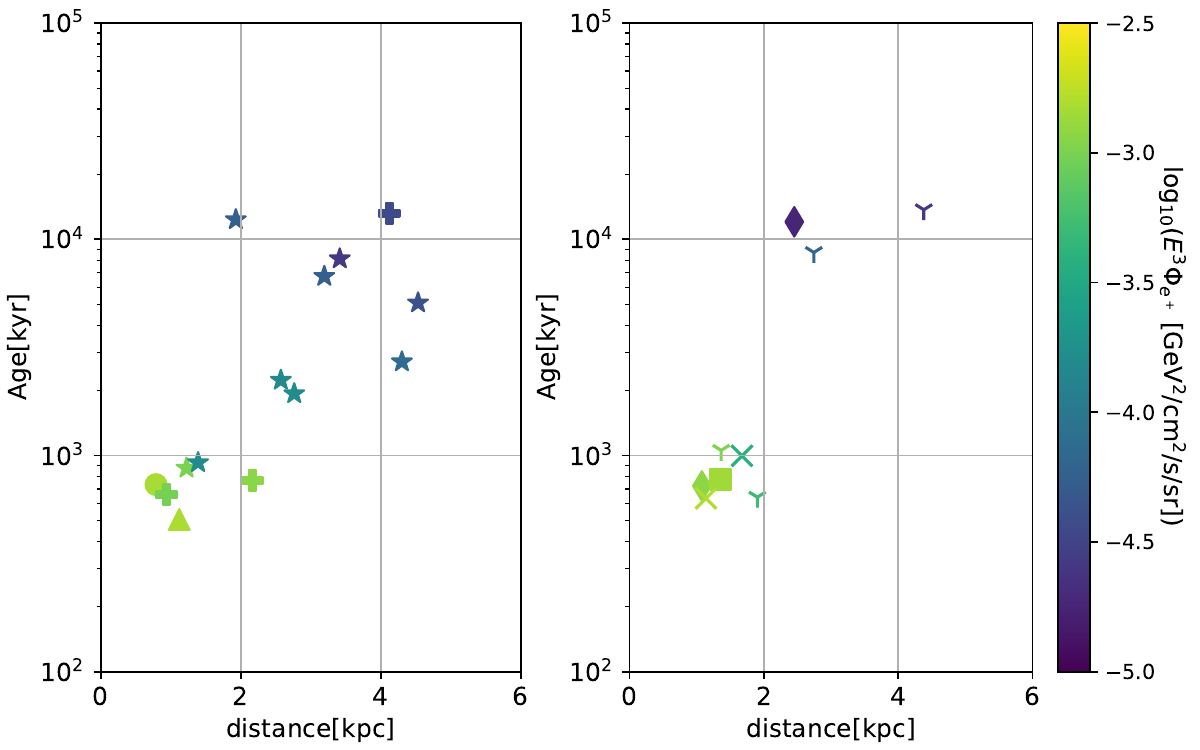}
        \caption{\texttt{ModA}, $\chi^2_{\rm red}<1.5$}
    \end{subfigure}%
    ~ 
    \begin{subfigure}[b]{0.5\textwidth}
        \centering
        \includegraphics[width=1\textwidth]{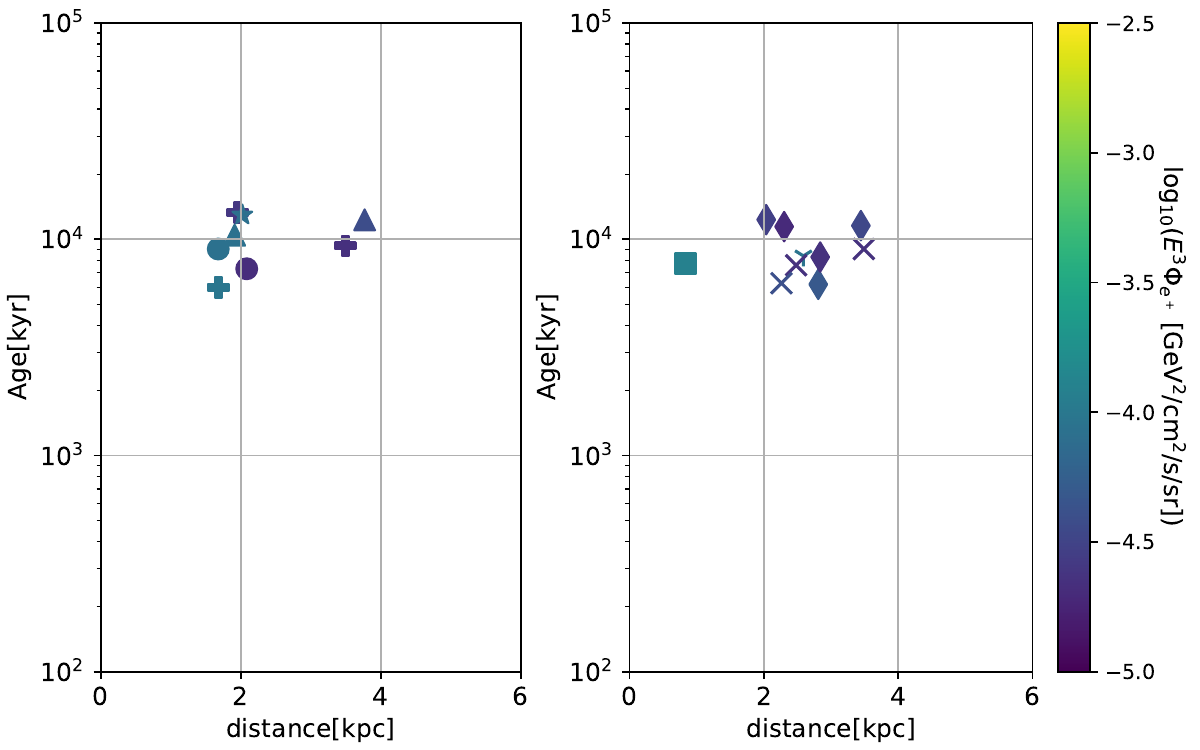}
        \caption{\texttt{ModA}, worst $\chi^2_{\rm red}$}
    \end{subfigure}
    \caption{Distance, age and maximum $E^3 \Phi_{e^+}(E)$ of the PWNe satisfying the {\em AMS-02 errors} criterion for \texttt{ModA} simulations with $\chi^2_{\rm red}<1.5$ (panel a) and with the worst $\chi^2_{\rm red}$  (panel b). In each panel, sources belonging to the same mock galaxy are reported with the same symbol, subdivided in two separate plots for better readability. The color scale is common to all  panels,
     and depicts the $\log_{10}$ of the maximum $e^+$ flux at the Earth in the AMS-02 data energy range. }
    \label{Fig R5}
\end{figure*}

In this section we scrutinize the physical properties of the simulated sources selected by the fit to be compatible with the AMS-02 data. 
For each Galactic realization of \texttt{ModA} with $\chi^2_{\rm red}<1.5$, we report in Figure~\ref{Fig R5}a the distance, age and maximum $E^3 \Phi_{e^+}(E)$ of the PWNe satisfying the {\em AMS-02 errors} criterion.
The data require 1 or 2 sources with high maximum $E^3 \Phi_{e^\pm}(E)$, with ages between 400 kyr and 2000 kyr and distances to the Earth less than 3 kpc. These sources produce fluxes peaked between 100 GeV and 500 GeV, allowing good explanation to the data. Fluxes from farther PWNe contribute less to the data. Sources with small maximum $E^3 \Phi_{e^\pm}(E)$ and with ages between 2000 kyr and $10^4$ kyr also satisfy the criterion, with flux peaks below 100 GeV where the secondaries are still the dominant component. We do not find any particular difference between all the simulation setups, 
except for \texttt{ModD} and \texttt{ModE}. As already noticed, since the {\it SLIM-MED} propagation implemented in \texttt{ModD} produces smoother fluxes, we find also some realizations with few more sources contributing with a bright flux to the $e^+$ data.

In Figure~\ref{Fig R5}b we report the distances, age and maximum $E^3 \Phi_{e^\pm}(E)$ values of the dominant PWNe for the mock galaxies with the worst $\chi^2$. These cases give best-fit to the data with the maximum values allowed by the priors for $A_S$ and the lowest values of $A_P$. 
Moreover, there are not sources which satisfy the {\em AMS-02 errors} criterion with an age between 400 kyr and 2000 kyr. In these galaxies, the trend of $E^3 \Phi_{e^\pm}(E)$ at high energies remains constant or decreases, and does not contribute sufficiently to the data above 50 GeV. 
To compensate this effect, the fit procedure demands the highest value of $A_S$.

\begin{figure*}[t]
    \centering
        \includegraphics[width=0.60\textwidth]{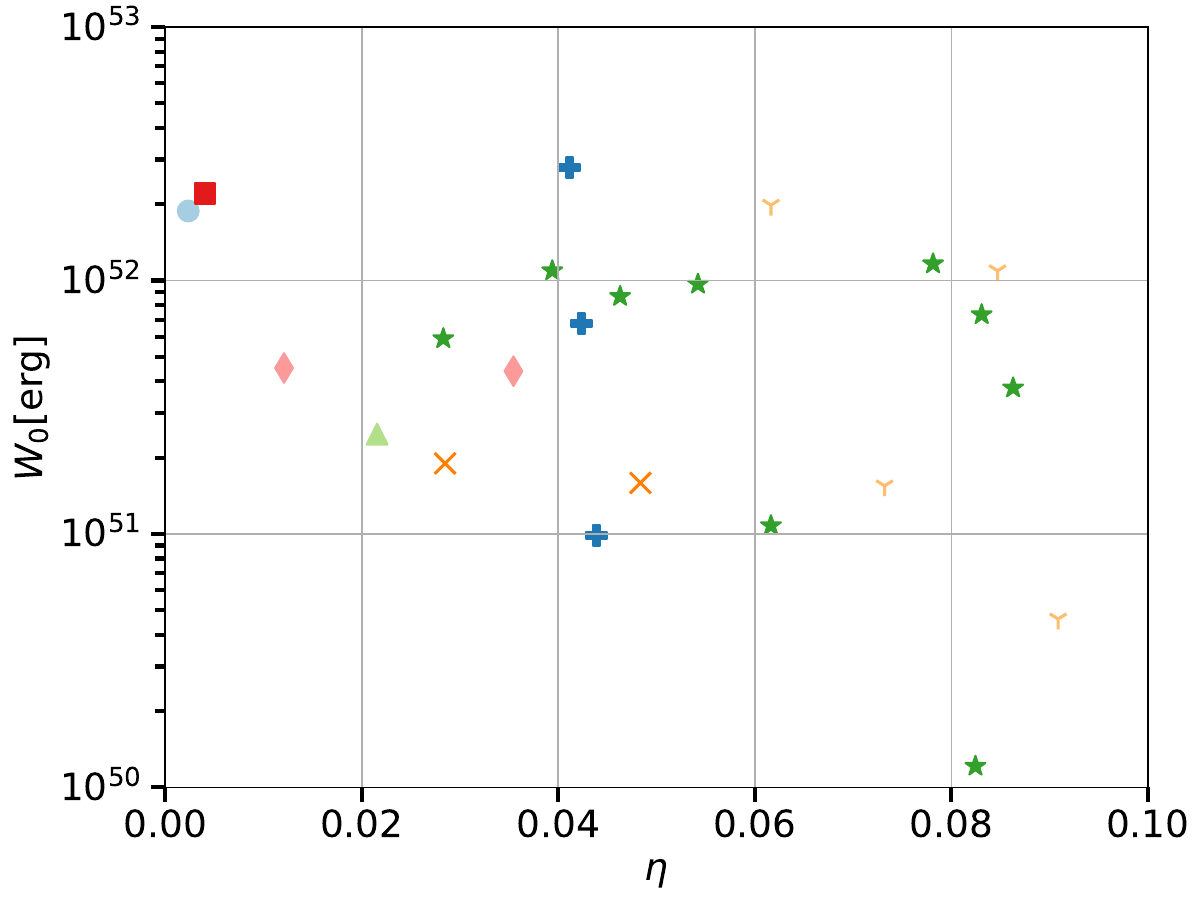}
    \caption{$W_0$ vs $\eta$ (see eqs.~\ref{eq:Erot} and \ref{eq:Etot}) values of the PWNe satisfying the {\em AMS-02 errors} criterion for \texttt{ModA} simulations with $\chi^2/d.o.f.<1.5$. Each galaxy is reported with the same  marker as in Figure~\ref{Fig R5}, with an assigned color.}
    \label{Fig R6}
\end{figure*}

In Figure~\ref{Fig R6} we report the distribution of best-fit efficiencies vs initial spin-down energy of each PWN that satisfies the 
{\em AMS-02 errors} criterion, for each simulation with $\chi^2_{\rm red}<1.5$.  The reported efficiencies are obtained
multiplying the simulated $\eta$ values associated to a single source with the $A_P$ obtained from the best fit of the corresponding galaxy. 
From Figure~\ref{Fig R6} it is evident that the efficiencies have a scattered distribution, and in most cases they have a value between 0.01 and 0.1, confirming the goodness of the $\eta$ interval initially chosen. Data hint at  a slight anti-correlation
between $\eta$ and $W_0$. In order to check that the characteristics of these pulsars are consistent with observations, we compute $\dot{E}$ from $W_0$ (see Section~\ref{sec:injection}), finding values quite common in nature. The ATNF catalog \cite{2005AJ....129.1993M} 
lists about 60 sources with $\dot{E}$ values higher than the maximum values obtained from sources reported in Figure~\ref{Fig R6},  namely $\dot{E}\sim 10^{36}$ erg s$^{-1}$. We do not directly compare the $W_0$ values, since for the sources of the ATNF catalog to compute $W_0$ we need to assume arbitrarily the value of $n$ and $\tau_0$. 
Instead, in our simulations we sample $n$ and we compute $\tau_0$ from the simulated parameters like $P_0$, that is also strictly connected to $W_0$.
For example, the cyan circle source of Figure~\ref{Fig R6} has $\dot{E} \sim 1.5  \times 10^{35}$ erg s$^{-1}$ and $\tau_0=0.7$ kyr.
Viceversa,  considering the J0002+6216 ATNF pulsar, with $\dot{E}= 1.5 \times 10^{35}$ erg s$^{-1}$ and $t = 306$ kyr, we obtain $W_0 \sim 4.7 \times 10^{49}$ erg for $\tau_0=10$ kyr and $n=3$, that changes to $6 \times 10^{50}$ erg for $\tau_0=0.7$ kyr and $n=3$,  and to $3.6 \times 10^{51}$ erg adopting $n=2.5$, spanning therefore different orders of magnitude.
We built our simulations starting from $P_0$ and $B$ distributions calibrated on observations. In the end, we obtain results which are consistent with these measurements.
We also outline that all the sources that satisfy the {\em AMS-02 errors} criterion have values of $\gamma_e$>1.7, for all the simulation setups. Lower values of $\gamma_e$ can produce peaked features incompatible with the smoothness of the AMS-02 data.
The adoption of a broken power-law injection spectrum $Q(E, t)$ (see  Section~\ref{sec:injection}) would probably limit the presence of peaked features at high energy, thanks to the soft spectral index above the break energy. We expect to find a slightly higher number of simulations compatible with the data, together with a slightly higher number of sources satisfying the {\em AMS-02 errors} criterion, producing negligible changes of our results.

\section{Conclusions}\label{sec:closing}
Despite the spatial and energetic distribution of pulsars and the details of the $e^\pm$ production, acceleration and release from these sources are not yet fully understood. The high-precision AMS-02 $e^+$ data are here used to constrain the main properties of the Galactic pulsar population and of the PWN acceleration needed to explain the observed CR flux. To this aim, we simulate a large number of  Galactic pulsar populations, calibrated on ATNF catalog observations.
Our simulations are conducted under different hypothesis about the pulsar spin-down and evolution properties, their radial distribution, and changing the propagation models for $e^\pm$ propagation in the Galaxy, following the most recent self-consistent modelings available in the literature. 
For each mock galaxy, we compute the $e^+$ flux at the Earth as the sum of the primary component due to PWNe emission, 
and a secondary component due to the fragmentation of CRs on the nuclei of the ISM. The result is fitted to AMS-02 $e^+$ data. 
 
Independently of the simulation scenario, we find that the vast majority of the galaxies realizations produce several wiggles in the total contribution and therefore they do not fit well the data. In all the tested setups, the number of mock galaxies with a $\chi^2_{\rm red}<1$ (2) does not exceed 1\% (4\%). 
The different features of the flux from single PWNs are due to the peculiar combination of the input parameters. We notice that the secondary flux, while decreasing with energy, practically forbids the realization of sharp cut-offs in the $e^+$ spectrum above TeV energies.
The galaxy realizations that fit properly the AMS-02 $e^+$ data have between 2-3 sources that produce a $e^+$ yield at the level of the data errors. Moreover, these pulsars provide a smooth spectrum that cover a wide energy range.
We find that the dominant contribution comes from sources located  between 1 and 3 kpc from the Earth. 
Despite the smaller effect from radiative cooling, the flux from sources within 1 kpc is lower due to the paucity of sources set along the spiral arms of the Galaxy. 
Sources dominating the observed spectrum have ages between 400 kyr and 2000 kyr and distances to the Earth less than 3 kpc. These sources produce fluxes peaked  between 100 GeV and 300 GeV, where AMS--02 data are the most constraining. 
Finally, we do not find any particular distribution for the pulsar efficiencies. In most cases they have a value between 0.01 and 0.1 consistently with what found in our previous papers \cite{DiMauro:2014iia,DiMauro:2019yvh}.

In ~\cite{Cholis_2018}, $e^{\pm}$ fluxes are computed from simulations of different combinations of pulsar spin-down properties, injection spectrum and propagation schemes. They consider sources younger than 10 Myr, relying on observations of pulsars with ages of order $10^5-10^7$ years. 
They constrain the space of pulsar and propagation models 
using the $e^+$ fraction and $e^++e^-$ data. Our simulations, while being more numerous, rely on distribution of parameters simulated according to complete pulsar population models. Our focus is on the characteristics of the pulsar population whose realization fits $e^+$ AMS-02 data. ~\cite{Cholis_2018} finds average values of $\eta$ in the range around 0.1-10\%, in agreement with our results for \texttt{ModA-B-C-D}.

Authors of ~\cite{Evoli_2021} adopt the bow-shock scenario explained in Section~\ref{sec:injection} and tested in our \texttt{ModE}, and analyze only a single pulsar population model.
They build mock galaxies sampling the $P_0$ value for each source from a gaussian distribution with $P_{0,mean}$ = 0.10 s and $P_{0,std}$ = 0.05 s obtained in \cite{Watters_2011}, a work based on young and energetic sources, producing on average pulsars more powerful with respect to our mock catalogs. They also adopt a higher pulsar birth rate (3/century) and they fix $\log_{10}(B)$ = 12.65 $\log_{10}({\rm G})$. Also, a different diffusion setup is considered, which implies a higher propagation scale length $\lambda$.
~\cite{Evoli_2021}  finds that the number of contributing sources to the $e^+$ flux is much larger($\sim 10^3$ at 1 TeV) than in this work and in previous works \cite{Fornieri_2020, Manconi:2020ipm, Recchia_2019} at all energies, due to the adopted diffusion model. We add that this result is probably also induced by the bow shock scenario and the spin down model adopted. With respect to their work 
we focus more on the characteristics that a pulsar realization has in order to fit the data. They find $\eta$ = 8.5\%, that rises to 42\% for $P_{0,mean}$ = 0.30 s and $P_{0,std}$ = 0.15 s, which is compatible with our \texttt{ModE}. 
However, more specific comparisons are difficult to perform, given the very different simulation setup between ~\cite{Evoli_2021} and this analysis.

With this paper we have shown the power of the AMS-02 CR $e^+$ data in constraining properties of PWNe supposed to be sources of $e^\pm$, in a measure to explain the flux data. Also for leptons, besides nuclei, an era has started for charged CRs to teach about
the Galaxy, in addition to the invaluable electromagnetic signals. 
\acknowledgments
We thank S. Recchia, M. Winkler and Y. Genolini for reading the manuscript and providing insightful comments.
SM thanks Y. Genolini and P. Mertsch for inspiring discussion.
 MDM research is supported by Fellini - Fellowship for Innovation at INFN, funded by the European Union’s Horizon 2020 research programme under the Marie Skłodowska-Curie Cofund Action, grant agreement no.~754496.
The work of FD and LO is  supported by 
the {\sc Departments of Excellence} grant awarded by the Italian Ministry of Education,
University and Research ({\sc Miur}), the 
Research grant {\sl The Dark Universe: A Synergic Multimessenger Approach}, No.
2017X7X85K funded by the {\sc Miur} and by the 
Research grant {\sc TAsP} (Theoretical Astroparticle Physics) funded by Istituto
Nazionale di Fisica Nucleare. 

\bibliographystyle{JHEP}
\bibliography{main}

\providecommand{\href}[2]{#2}\begingroup\raggedright\begin{thebibliography}{10}

\bibitem{Adriani:2013uda}
{\scshape PAMELA} collaboration, \emph{{Cosmic-Ray Positron Energy Spectrum
  Measured by PAMELA}},
  \href{https://doi.org/10.1103/PhysRevLett.111.081102}{\emph{Phys. Rev. Lett.}
  {\bfseries 111} (2013) 081102}
  [\href{https://arxiv.org/abs/1308.0133}{{\ttfamily 1308.0133}}].

\bibitem{Ackermann_2012}
M.~Ackermann, M.~Ajello, A.~Allafort, W.B.~Atwood, L.~Baldini, G.~Barbiellini
  et~al., \emph{Measurement of separate cosmic-ray electron and positron
  spectra with the fermi large area telescope},
  \href{https://doi.org/10.1103/physrevlett.108.011103}{\emph{Physical Review
  Letters} {\bfseries 108} (2012) }.

\bibitem{PhysRevLett.122.041102}
{\scshape AMS Collaboration} collaboration, \emph{Towards understanding the
  origin of cosmic-ray positrons},
  \href{https://doi.org/10.1103/PhysRevLett.122.041102}{\emph{Phys. Rev. Lett.}
  {\bfseries 122} (2019) 041102}.

\bibitem{Delahaye:2008ua}
T.~Delahaye, F.~Donato, N.~Fornengo, J.~Lavalle, R.~Lineros, P.~Salati et~al.,
  \emph{{Galactic secondary positron flux at the Earth}},
  \href{https://doi.org/10.1051/0004-6361/200811130}{\emph{Astron. Astrophys.}
  {\bfseries 501} (2009) 821}
  [\href{https://arxiv.org/abs/0809.5268}{{\ttfamily 0809.5268}}].

\bibitem{Diesing:2020jtm}
R.~Diesing and D.~Caprioli, \emph{{Nonsecondary origin of cosmic ray
  positrons}}, \href{https://doi.org/10.1103/PhysRevD.101.103030}{\emph{Phys.
  Rev. D} {\bfseries 101} (2020) 103030}
  [\href{https://arxiv.org/abs/2001.02240}{{\ttfamily 2001.02240}}].

\bibitem{Mlyshev:2009twa}
D.~Malyshev, I.~Cholis and J.~Gelfand, \emph{{Pulsars versus Dark Matter
  Interpretation of ATIC/PAMELA}},
  \href{https://doi.org/10.1103/PhysRevD.80.063005}{\emph{Phys. Rev.}
  {\bfseries D80} (2009) 063005}
  [\href{https://arxiv.org/abs/0903.1310}{{\ttfamily 0903.1310}}].

\bibitem{2009_Grasso}
D.~Grasso, S.~Profumo, A.~Strong, L.~Baldini, R.~Bellazzini, E.~Bloom et~al.,
  \emph{On possible interpretations of the high energy electron–positron
  spectrum measured by the fermi large area telescope},
  \href{https://doi.org/10.1016/j.astropartphys.2009.07.003}{\emph{Astroparticle
  Physics} {\bfseries 32} (2009) 140–151}.

\bibitem{2009_Hooper}
D.~Hooper, P.~Blasi and P.D.~Serpico, \emph{Pulsars as the sources of high
  energy cosmic ray positrons},
  \href{https://doi.org/10.1088/1475-7516/2009/01/025}{\emph{Journal of
  Cosmology and Astroparticle Physics} {\bfseries 2009} (2009) 025–025}.

\bibitem{2010_Kawanaka}
N.~Kawanaka, K.~Ioka and M.M.~Nojiri, \emph{Is cosmic ray electron excess from
  pulsars spiky or smooth?: Continuous and multiple electron/positron
  injections}, \href{https://doi.org/10.1088/0004-637x/710/2/958}{\emph{The
  Astrophysical Journal} {\bfseries 710} (2010) 958–963}.

\bibitem{2011_Kashiyama}
K.~Kashiyama, K.~Ioka and N.~Kawanaka, \emph{White dwarf pulsars as possible
  cosmic ray electron-positron factories},
  \href{https://doi.org/10.1103/physrevd.83.023002}{\emph{Physical Review D}
  {\bfseries 83} (2011) }.

\bibitem{2012_Kisaka}
S.~Kisaka and N.~Kawanaka, \emph{Tev cosmic-ray electrons from millisecond
  pulsars},
  \href{https://doi.org/10.1111/j.1365-2966.2012.20576.x}{\emph{Monthly Notices
  of the Royal Astronomical Society} {\bfseries 421} (2012) 3543–3549}.

\bibitem{Gaggero:2013nfa}
D.~Gaggero, L.~Maccione, D.~Grasso, G.~Di~Bernardo and C.~Evoli, \emph{{PAMELA
  and AMS-02 $e^+$ and $e^-$ spectra are reproduced by three-dimensional
  cosmic-ray modeling}},
  \href{https://doi.org/10.1103/PhysRevD.89.083007}{\emph{Phys. Rev.}
  {\bfseries D89} (2014) 083007}
  [\href{https://arxiv.org/abs/1311.5575}{{\ttfamily 1311.5575}}].

\bibitem{2013_cholis}
I.~Cholis and D.~Hooper, \emph{Dark matter and pulsar origins of the rising
  cosmic ray positron fraction in light of new data from the ams},
  \href{https://doi.org/10.1103/physrevd.88.023013}{\emph{Physical Review D}
  {\bfseries 88} (2013) }.

\bibitem{2014JCAP...04..006D}
M.~{Di Mauro}, F.~{Donato}, N.~{Fornengo}, R.~{Lineros} and A.~{Vittino},
  \emph{{Interpretation of AMS-02 electrons and positrons data}},
  \href{https://doi.org/10.1088/1475-7516/2014/04/006}{\emph{\jcap} {\bfseries
  4} (2014) 6} [\href{https://arxiv.org/abs/1402.0321}{{\ttfamily 1402.0321}}].

\bibitem{Boudaud:2014dta}
M.~Boudaud et~al., \emph{{A new look at the cosmic ray positron fraction}},
  \href{https://doi.org/10.1051/0004-6361/201425197}{\emph{Astron. Astrophys.}
  {\bfseries 575} (2015) A67}
  [\href{https://arxiv.org/abs/1410.3799}{{\ttfamily 1410.3799}}].

\bibitem{Tomassetti:2015cva}
N.~Tomassetti and F.~Donato, \emph{{The Connection Between the Positron
  Fraction Anomaly and the Spectral Features in Galactic Cosmic-Ray Hadrons}},
  \href{https://doi.org/10.1088/2041-8205/803/2/L15}{\emph{Astrophys. J.}
  {\bfseries 803} (2015) L15}
  [\href{https://arxiv.org/abs/1502.06150}{{\ttfamily 1502.06150}}].

\bibitem{Lipari:2018usj}
P.~Lipari, \emph{{Spectral shapes of the fluxes of electrons and positrons and
  the average residence time of cosmic rays in the Galaxy}},
  \href{https://doi.org/10.1103/PhysRevD.99.043005}{\emph{Phys. Rev.}
  {\bfseries D99} (2019) 043005}
  [\href{https://arxiv.org/abs/1810.03195}{{\ttfamily 1810.03195}}].

\bibitem{Cholis_2018}
I.~Cholis, T.~Karwal and M.~Kamionkowski, \emph{Studying the milky way pulsar
  population with cosmic-ray leptons},
  \href{https://doi.org/10.1103/physrevd.98.063008}{\emph{Physical Review D}
  {\bfseries 98} (2018) }.

\bibitem{2016JCAP...05..031D}
M.~{Di Mauro}, F.~{Donato}, N.~{Fornengo} and A.~{Vittino}, \emph{{Dark matter
  vs. astrophysics in the interpretation of AMS-02 electron and positron
  data}}, \href{https://doi.org/10.1088/1475-7516/2016/05/031}{\emph{\jcap}
  {\bfseries 5} (2016) 031} [\href{https://arxiv.org/abs/1507.07001}{{\ttfamily
  1507.07001}}].

\bibitem{Mertsch:2020dcy}
P.~Mertsch, A.~Vittino and S.~Sarkar, \emph{{Explaining cosmic ray antimatter
  with secondaries from old supernova remnants}},
  \href{https://arxiv.org/abs/2012.12853}{{\ttfamily 2012.12853}}.

\bibitem{Bykov:2017xpo}
A.M.~Bykov, E.~Amato, A.E.~Petrov, A.M.~Krassilchtchikov and K.P.~Levenfish,
  \emph{{Pulsar wind nebulae with bow shocks: non-thermal radiation and cosmic
  ray leptons}}, \href{https://doi.org/10.1007/s11214-017-0371-7}{\emph{Space
  Sci. Rev.} {\bfseries 207} (2017) 235}
  [\href{https://arxiv.org/abs/1705.00950}{{\ttfamily 1705.00950}}].

\bibitem{Amato:2020zfv}
E.~Amato, \emph{{The theory of Pulsar Wind Nebulae: recent progress}},
  \href{https://doi.org/10.22323/1.354.0033}{\emph{PoS} {\bfseries HEPROVII}
  (2020) 033} [\href{https://arxiv.org/abs/2001.04442}{{\ttfamily
  2001.04442}}].

\bibitem{Gaensler_2003}
B.M.~Gaensler, N.S.~Schulz, V.M.~Kaspi, M.J.~Pivovaroff and W.E.~Becker,
  \emph{{XMM}-{NewtonObservations} of {PSR} b1823-13: An asymmetric synchrotron
  nebula around a vela-like pulsar},
  \href{https://doi.org/10.1086/368356}{\emph{The Astrophysical Journal}
  {\bfseries 588} (2003) 441}.

\bibitem{2017hsn..book.2159S}
P.~{Slane}, \emph{{Pulsar Wind Nebulae, Handbook of Supernovae, ISBN
  978-3-319-21845-8.~Springer International Publishing AG, 2017, p.~2159}},
  \href{https://arxiv.org/abs/1703.09311}{{\ttfamily 1703.09311}}.

\bibitem{2009ApJ...700L.127A}
A.A.~{Abdo}, B.T.~{Allen}, T.~{Aune} et~al., \emph{{Milagro Observations of
  Multi-TeV Emission from Galactic Sources in the Fermi Bright Source List}},
  \href{https://doi.org/10.1088/0004-637X/700/2/L127}{\emph{\apjl} {\bfseries
  700} (2009) L127} [\href{https://arxiv.org/abs/0904.1018}{{\ttfamily
  0904.1018}}].

\bibitem{Abeysekara:2017science}
{\scshape HAWC} collaboration, \emph{{Extended gamma-ray sources around pulsars
  constrain the origin of the positron flux at Earth}},
  \href{https://doi.org/10.1126/science.aan4880}{\emph{Science} {\bfseries 358}
  (2017) 911} [\href{https://arxiv.org/abs/1711.06223}{{\ttfamily
  1711.06223}}].

\bibitem{HAWC:2019tcx}
{\scshape HAWC} collaboration, \emph{{Multiple Galactic Sources with Emission
  Above 56 TeV Detected by HAWC}},
  \href{https://doi.org/10.1103/PhysRevLett.124.021102}{\emph{Phys. Rev. Lett.}
  {\bfseries 124} (2020) 021102}
  [\href{https://arxiv.org/abs/1909.08609}{{\ttfamily 1909.08609}}].

\bibitem{LHAASO:2021crt}
{\scshape LHAASO} collaboration, \emph{{Extended Very-High-Energy Gamma-Ray
  Emission Surrounding PSR J0622+3749 Observed by LHAASO-KM2A}},
  \href{https://doi.org/10.1103/PhysRevLett.126.241103}{\emph{Phys. Rev. Lett.}
  {\bfseries 126} (2021) 241103}
  [\href{https://arxiv.org/abs/2106.09396}{{\ttfamily 2106.09396}}].

\bibitem{DiMauro:2019yvh}
M.~Di~Mauro, S.~Manconi and F.~Donato, \emph{{Detection of a $\gamma$-ray halo
  around Geminga with the Fermi -LAT data and implications for the positron
  flux}}, \href{https://doi.org/10.1103/PhysRevD.100.123015}{\emph{Phys. Rev.
  D} {\bfseries 100} (2019) 123015}
  [\href{https://arxiv.org/abs/1903.05647}{{\ttfamily 1903.05647}}].

\bibitem{Linden:2017vvb}
T.~Linden, K.~Auchettl, J.~Bramante, I.~Cholis, K.~Fang, D.~Hooper et~al.,
  \emph{{Using HAWC to discover invisible pulsars}},
  \href{https://doi.org/10.1103/PhysRevD.96.103016}{\emph{Phys. Rev.}
  {\bfseries D96} (2017) 103016}
  [\href{https://arxiv.org/abs/1703.09704}{{\ttfamily 1703.09704}}].

\bibitem{DiMauro:2019hwn}
M.~Di~Mauro, S.~Manconi and F.~Donato, \emph{{Evidences of low-diffusion
  bubbles around Galactic pulsars}},
  \href{https://doi.org/10.1103/PhysRevD.101.103035}{\emph{Phys. Rev. D}
  {\bfseries 101} (2020) 103035}
  [\href{https://arxiv.org/abs/1908.03216}{{\ttfamily 1908.03216}}].

\bibitem{DiMauro:2020jbz}
M.~Di~Mauro, S.~Manconi, M.~Negro and F.~Donato, \emph{{Investigating
  $\gamma$-ray halos around three HAWC bright sources in Fermi-LAT data}},
  \href{https://arxiv.org/abs/2012.05932}{{\ttfamily 2012.05932}}.

\bibitem{Evoli:2018aza}
C.~Evoli, T.~Linden and G.~Morlino, \emph{{Self-generated cosmic-ray
  confinement in TeV halos: Implications for TeV $\gamma$-ray emission and the
  positron excess}},
  \href{https://doi.org/10.1103/PhysRevD.98.063017}{\emph{Phys. Rev.}
  {\bfseries D98} (2018) 063017}
  [\href{https://arxiv.org/abs/1807.09263}{{\ttfamily 1807.09263}}].

\bibitem{Lopez-Coto:2017pbk}
R.~L\'opez-Coto and G.~Giacinti, \emph{{Constraining the properties of the
  magnetic turbulence in the Geminga region using HAWC $\gamma$-ray data}},
  \href{https://doi.org/10.1093/mnras/sty1821}{\emph{Mon. Not. Roy. Astron.
  Soc.} {\bfseries 479} (2018) 4526}
  [\href{https://arxiv.org/abs/1712.04373}{{\ttfamily 1712.04373}}].

\bibitem{Liu:2019zyj}
R.-Y.~Liu, H.~Yan and H.~Zhang, \emph{{Understanding the Multiwavelength
  Observation of Geminga\textquoteright{}s Tev Halo: The Role of Anisotropic
  Diffusion of Particles}},
  \href{https://doi.org/10.1103/PhysRevLett.123.221103}{\emph{Phys. Rev. Lett.}
  {\bfseries 123} (2019) 221103}
  [\href{https://arxiv.org/abs/1904.11536}{{\ttfamily 1904.11536}}].

\bibitem{Fang:2019iym}
K.~Fang, X.-J.~Bi and P.-F.~Yin, \emph{{Possible origin of the slow-diffusion
  region around Geminga}},
  \href{https://doi.org/10.1093/mnras/stz1974}{\emph{Mon. Not. Roy. Astron.
  Soc.} {\bfseries 488} (2019) 4074}
  [\href{https://arxiv.org/abs/1903.06421}{{\ttfamily 1903.06421}}].

\bibitem{Recchia:2021kty}
S.~Recchia, M.~Di~Mauro, F.A.~Aharonian, F.~Donato, S.~Gabici and S.~Manconi,
  \emph{{Does the Geminga $\gamma$-ray halo imply slow diffusion around
  pulsars?}},  \href{https://arxiv.org/abs/2106.02275}{{\ttfamily 2106.02275}}.

\bibitem{Manconi:2016byt}
S.~Manconi, M.~Di~Mauro and F.~Donato, \emph{{Dipole anisotropy in cosmic
  electrons and positrons: inspection on local sources}},
  \href{https://doi.org/10.1088/1475-7516/2017/01/006}{\emph{JCAP} {\bfseries
  01} (2017) 006} [\href{https://arxiv.org/abs/1611.06237}{{\ttfamily
  1611.06237}}].

\bibitem{Fornieri_2020}
O.~Fornieri, D.~Gaggero and D.~Grasso, \emph{Features in cosmic-ray lepton data
  unveil the properties of nearby cosmic accelerators},
  \href{https://doi.org/10.1088/1475-7516/2020/02/009}{\emph{Journal of
  Cosmology and Astroparticle Physics} {\bfseries 2020} (2020) 009}.

\bibitem{DiMauro:2020cbn}
M.~Di~Mauro, F.~Donato and S.~Manconi, \emph{{On the interpretation of the
  latest AMS-02 cosmic ray electron spectrum}},
  \href{https://arxiv.org/abs/2010.13825}{{\ttfamily 2010.13825}}.

\bibitem{Evoli_2021}
C.~Evoli, E.~Amato, P.~Blasi and R.~Aloisio, \emph{Galactic factories of
  cosmic-ray electrons and positrons},
  \href{https://doi.org/10.1103/physrevd.103.083010}{\emph{Physical Review D}
  {\bfseries 103} (2021) }.

\bibitem{Manconi:2020ipm}
S.~Manconi, M.~Di~Mauro and F.~Donato, \emph{{Contribution of pulsars to
  cosmic-ray positrons in light of recent observation of inverse-Compton
  halos}}, \href{https://doi.org/10.1103/PhysRevD.102.023015}{\emph{Phys. Rev.
  D} {\bfseries 102} (2020) 023015}
  [\href{https://arxiv.org/abs/2001.09985}{{\ttfamily 2001.09985}}].

\bibitem{Lorimer_2006}
D.R.~Lorimer, A.J.~Faulkner, A.G.~Lyne, R.N.~Manchester, M.~Kramer,
  M.A.~McLaughlin et~al., \emph{The parkes multibeam pulsar survey - vi.
  discovery and timing of 142 pulsars and a galactic population analysis},
  \href{https://doi.org/10.1111/j.1365-2966.2006.10887.x}{\emph{Monthly Notices
  of the Royal Astronomical Society} {\bfseries 372} (2006) 777–800}.

\bibitem{Chakraborty:2020lbu}
A.~Chakraborty and M.~Bagchi, \emph{{Understanding the Galactic population of
  normal pulsars: A leap forward}},
  \href{https://arxiv.org/abs/2012.13243}{{\ttfamily 2012.13243}}.

\bibitem{2019_Manconi}
S.~Manconi, M.D.~Mauro and F.~Donato, \emph{Multi-messenger constraints to the
  local emission of cosmic-ray electrons},
  \href{https://doi.org/10.1088/1475-7516/2019/04/024}{\emph{Journal of
  Cosmology and Astroparticle Physics} {\bfseries 2019} (2019) 024–024}.

\bibitem{Abdollahi:2017nat}
{\scshape Fermi-LAT} collaboration, \emph{{Cosmic-ray electron-positron
  spectrum from 7 GeV to 2 TeV with the Fermi Large Area Telescope}},
  \href{https://doi.org/10.1103/PhysRevD.95.082007}{\emph{Phys. Rev. D}
  {\bfseries 95} (2017) 082007}
  [\href{https://arxiv.org/abs/1704.07195}{{\ttfamily 1704.07195}}].

\bibitem{PhysRevLett.122.101101}
{\scshape AMS} collaboration, \emph{Towards understanding the origin of
  cosmic-ray electrons},
  \href{https://doi.org/10.1103/PhysRevLett.122.101101}{\emph{Phys. Rev. Lett.}
  {\bfseries 122} (2019) 101101}.

\bibitem{2017_DAMPE}
{\scshape DAMPE} collaboration, \emph{Direct detection of a break in the
  teraelectronvolt cosmic-ray spectrum of electrons and positrons},
  \href{https://doi.org/10.1038/nature24475}{\emph{Nature} {\bfseries 552}
  (2017) 63–66}.

\bibitem{PhysRevLett.120.261102}
{\scshape CALET} collaboration, \emph{Extended measurement of the cosmic-ray
  electron and positron spectrum from 11 gev to 4.8 tev with the calorimetric
  electron telescope on the international space station},
  \href{https://doi.org/10.1103/PhysRevLett.120.261102}{\emph{Phys. Rev. Lett.}
  {\bfseries 120} (2018) 261102}.

\bibitem{Abeysekara:2017hyn}
A.U.~Abeysekara et~al., \emph{{The 2HWC HAWC Observatory Gamma Ray Catalog}},
  \href{https://doi.org/10.3847/1538-4357/aa7556}{\emph{Astrophys. J.}
  {\bfseries 843} (2017) 40}
  [\href{https://arxiv.org/abs/1702.02992}{{\ttfamily 1702.02992}}].

\bibitem{Yuksel:2008rf}
H.~Yuksel, M.D.~Kistler and T.~Stanev, \emph{{TeV Gamma Rays from Geminga and
  the Origin of the GeV Positron Excess}},
  \href{https://doi.org/10.1103/PhysRevLett.103.051101}{\emph{Phys. Rev. Lett.}
  {\bfseries 103} (2009) 051101}
  [\href{https://arxiv.org/abs/0810.2784}{{\ttfamily 0810.2784}}].

\bibitem{Torres:2014iua}
D.F.~Torres, A.~Cillis, J.~Martín and E.~de~Oña~Wilhelmi,
  \emph{{Time-dependent modeling of TeV-detected, young pulsar wind nebulae}},
  \href{https://doi.org/10.1016/j.jheap.2014.02.001}{\emph{JHEAp} {\bfseries
  1-2} (2014) 31} [\href{https://arxiv.org/abs/1402.5485}{{\ttfamily
  1402.5485}}].

\bibitem{Sushch:2013tna}
I.~Sushch and B.~Hnatyk, \emph{{Modelling of the radio emission from the Vela
  supernova remnant}},
  \href{https://doi.org/10.1051/0004-6361/201322569}{\emph{Astron. Astrophys.}
  {\bfseries 561} (2014) A139}
  [\href{https://arxiv.org/abs/1312.0777}{{\ttfamily 1312.0777}}].

\bibitem{Buesching:2008hr}
I.~Buesching, O.C.~de~Jager, M.S.~Potgieter and C.~Venter, \emph{{A Cosmic Ray
  Positron Anisotropy due to Two Middle-Aged, Nearby Pulsars?}},
  \href{https://doi.org/10.1086/588465}{\emph{Astrophys. J.} {\bfseries 678}
  (2008) L39} [\href{https://arxiv.org/abs/0804.0220}{{\ttfamily 0804.0220}}].

\bibitem{Ridley_2010}
J.P.~Ridley and D.R.~Lorimer, \emph{Isolated pulsar spin evolution on the
  diagram},
  \href{https://doi.org/10.1111/j.1365-2966.2010.16342.x}{\emph{Monthly Notices
  of the Royal Astronomical Society} {\bfseries 404} (2010) 1081–1088}.

\bibitem{2011ASSP...21..624B}
P.~{Blasi} and E.~{Amato}, \emph{{Positrons from pulsar winds}},
  {\emph{Astrophysics and Space Science Proceedings} {\bfseries 21} (2011) 624}
  [\href{https://arxiv.org/abs/1007.4745}{{\ttfamily 1007.4745}}].

\bibitem{van_der_Swaluw_2003}
E.~van~der Swaluw, A.~Achterberg, Y.A.~Gallant, T.P.~Downes and R.~Keppens,
  \emph{Interaction of high-velocity pulsars with supernova remnant shells},
  \href{https://doi.org/10.1051/0004-6361:20021488}{\emph{\aap} {\bfseries 397}
  (2003) 913–920}.

\bibitem{2010A&A...524A..51D}
T.~{Delahaye}, J.~{Lavalle}, R.~{Lineros}, F.~{Donato} and N.~{Fornengo},
  \emph{{Galactic electrons and positrons at the Earth: new estimate of the
  primary and secondary fluxes}},
  \href{https://doi.org/10.1051/0004-6361/201014225}{\emph{\aap} {\bfseries
  524} (2010) A51} [\href{https://arxiv.org/abs/1002.1910}{{\ttfamily
  1002.1910}}].

\bibitem{Manconi:2018azw}
S.~Manconi, M.~Di~Mauro and F.~Donato, \emph{{Multi-messenger constraints to
  the local emission of cosmic-ray electrons}},
  \href{https://doi.org/10.1088/1475-7516/2019/04/024}{\emph{JCAP} {\bfseries
  04} (2019) 024} [\href{https://arxiv.org/abs/1803.01009}{{\ttfamily
  1803.01009}}].

\bibitem{dimauro2021multimessenger}
M.D.~Mauro and M.W.~Winkler, \emph{Multimessenger constraints on the dark
  matter interpretation of the fermi-lat galactic center excess},
  \href{https://arxiv.org/abs/2101.11027}{{\ttfamily 2101.11027}}.

\bibitem{Weinrich:2020ftb}
N.~Weinrich, M.~Boudaud, L.~Derome, Y.~Genolini, J.~Lavalle, D.~Maurin et~al.,
  \emph{{Galactic halo size in the light of recent AMS-02 data}},
  \href{https://doi.org/10.1051/0004-6361/202038064}{\emph{Astron. Astrophys.}
  {\bfseries 639} (2020) A74}
  [\href{https://arxiv.org/abs/2004.00441}{{\ttfamily 2004.00441}}].

\bibitem{Vernetto:2016alq}
S.~Vernetto and P.~Lipari, \emph{{Absorption of very high energy gamma rays in
  the Milky Way}},
  \href{https://doi.org/10.1103/PhysRevD.94.063009}{\emph{Phys. Rev.}
  {\bfseries D94} (2016) 063009}
  [\href{https://arxiv.org/abs/1608.01587}{{\ttfamily 1608.01587}}].

\bibitem{genolini2021new}
Y.~Génolini, M.~Boudaud, M.~Cirelli, L.~Derome, J.~Lavalle, D.~Maurin et~al.,
  \emph{New minimal, median, and maximal propagation models for dark matter
  searches with galactic cosmic rays},
  \href{https://arxiv.org/abs/2103.04108}{{\ttfamily 2103.04108}}.

\bibitem{Recchia_2019}
S.~Recchia, S.~Gabici, F.~Aharonian and J.~Vink, \emph{Local fading accelerator
  and the origin of tev cosmic ray electrons},
  \href{https://doi.org/10.1103/physrevd.99.103022}{\emph{Physical Review D}
  {\bfseries 99} (2019) }.

\bibitem{2006ApJ...643..332F}
C.-A.~{Faucher-Giguere} and V.M.~{Kaspi}, \emph{{Birth and Evolution of
  Isolated Radio Pulsars}}, \href{https://doi.org/10.1086/501516}{\emph{\apj}
  {\bfseries 643} (2006) 332}
  [\href{https://arxiv.org/abs/astro-ph/0512585}{{\ttfamily
  astro-ph/0512585}}].

\bibitem{2004IAUS..218..105L}
D.R.~{Lorimer}, \emph{{The Galactic Population and Birth Rate of Radio
  Pulsars}},  \href{https://arxiv.org/abs/astro-ph/0308501}{{\ttfamily
  astro-ph/0308501}}.

\bibitem{Keane:2008jj}
E.F.~Keane and M.~Kramer, \emph{{On the birthrates of Galactic neutron stars}},
  \href{https://doi.org/10.1111/j.1365-2966.2008.14045.x}{\emph{Mon. Not. Roy.
  Astron. Soc.} {\bfseries 391} (2008) 2009}
  [\href{https://arxiv.org/abs/0810.1512}{{\ttfamily 0810.1512}}].

\bibitem{Nigro_2019}
C.~Nigro, C.~Deil, R.~Zanin, T.~Hassan, J.~King, J.E.~Ruiz et~al.,
  \emph{Towards open and reproducible multi-instrument analysis in gamma-ray
  astronomy},
  \href{https://doi.org/10.1051/0004-6361/201834938}{\emph{Astronomy and
  Astrophysics} {\bfseries 625} (2019) A10}.

\bibitem{CTAConsortium:2017xaq}
{\scshape CTA Consortium} collaboration, \emph{{Gammapy - A prototype for the
  CTA science tools}}, \href{https://doi.org/10.22323/1.301.0766}{\emph{PoS}
  {\bfseries ICRC2017} (2018) 766}
  [\href{https://arxiv.org/abs/1709.01751}{{\ttfamily 1709.01751}}].

\bibitem{LATTIMER_2007}
J.~Lattimer and M.~Prakash, \emph{Neutron star observations: Prognosis for
  equation of state constraints},
  \href{https://doi.org/10.1016/j.physrep.2007.02.003}{\emph{Physics Reports}
  {\bfseries 442} (2007) 109–165}.

\bibitem{Gaensler:2006ua}
B.M.~Gaensler and P.O.~Slane, \emph{{The evolution and structure of pulsar wind
  nebulae}},
  \href{https://doi.org/10.1146/annurev.astro.44.051905.092528}{\emph{Ann. Rev.
  Astron. Astrophys.} {\bfseries 44} (2006) 17}
  [\href{https://arxiv.org/abs/astro-ph/0601081}{{\ttfamily
  astro-ph/0601081}}].

\bibitem{2005AJ....129.1993M}
R.N.~{Manchester}, G.B.~{Hobbs}, A.~{Teoh} and M.~{Hobbs}, \emph{{The Australia
  Telescope National Facility Pulsar Catalogue}},
  \href{https://doi.org/10.1086/428488}{\emph{\aj} {\bfseries 129} (2005) 1993}
  [\href{https://arxiv.org/abs/astro-ph/0412641}{{\ttfamily
  astro-ph/0412641}}].

\bibitem{Case_1998}
G.L.~Case and D.~Bhattacharya, \emph{A new sigma‐d relation and its
  application to the galactic supernova remnant distribution},
  \href{https://doi.org/10.1086/306089}{\emph{The Astrophysical Journal}
  {\bfseries 504} (1998) 761–772}.

\bibitem{Yusifov_2004}
I.~Yusifov and I.~Küçük, \emph{Revisiting the radial distribution of pulsars
  in the galaxy}, \href{https://doi.org/10.1051/0004-6361:20040152}{\emph{\aap}
  {\bfseries 422} (2004) 545–553}.

\bibitem{Boudaud:2016jvj}
M.~Boudaud, E.F.~Bueno, S.~Caroff, Y.~Genolini, V.~Poulin, V.~Poireau et~al.,
  \emph{{The pinching method for Galactic cosmic ray positrons: implications in
  the light of precision measurements}},
  \href{https://doi.org/10.1051/0004-6361/201630321}{\emph{Astron. Astrophys.}
  {\bfseries 605} (2017) A17}
  [\href{https://arxiv.org/abs/1612.03924}{{\ttfamily 1612.03924}}].

\bibitem{article_Chang}
J.~Chang et~al., \emph{An excess of cosmic ray electrons at energies of
  300–800 gev}, \href{https://doi.org/10.1038/nature07477}{\emph{Nature}
  {\bfseries 456} (2008) 362}.

\bibitem{torii2008highenergy}
S.~Torii et~al., \emph{High-energy electron observations by ppb-bets flight in
  antarctica},  2008.

\bibitem{DiMauro:2014iia}
M.~Di~Mauro, F.~Donato, N.~Fornengo et~al., \emph{{Interpretation of AMS-02
  electrons and positrons data}},
  \href{https://doi.org/10.1088/1475-7516/2014/04/006}{\emph{\jcap} {\bfseries
  1404} (2014) 006} [\href{https://arxiv.org/abs/1402.0321}{{\ttfamily
  1402.0321}}].

\bibitem{Watters_2011}
K.P.~Watters and R.W.~Romani, \emph{The galactic population $\gamma$-ray
  pulsars}, \href{https://doi.org/10.1088/0004-637x/727/2/123}{\emph{The
  Astrophysical Journal} {\bfseries 727} (2011) 123}.

\end{thebibliography}\endgroup

\end{document}